\def\e{{\textrm e}}
\def\del{\partial}
\def\half{{1\over2}}

\def\abs#1{{\left|{#1}\right|}}
\def\vev#1{\langle #1 \rangle}
\def\del{\partial}
\def\half{{1\over2}}

\def\abs#1{{\left|{#1}\right|}}
\def\vev#1{\langle #1 \rangle}

\def\del{\partial}
\def\dslash{\del\kern-0.55em\raise 0.14ex\hbox{/}}

\def\rough#1{\raise.3ex\hbox{$#1$\kern-.75em\lower1ex\hbox{$\sim$}}}

\newcommand{\PRD}[3]{{\it Phys. Rev.} {\bf D{#1}} (19{#3}) {#2}}

\newcommand{\PRDM}[3]{{\it Phys. Rev.} {\bf D{#1}} {#2} (20{#3})}

\newcommand{\NPB}[3]{{\it Nucl. Phys.} {\bf B{#1}} (19{#2}) {#3}}
\newcommand{\NPBM}[3]{{\it Nucl. Phys.} {\bf B{#1}} (20{#2}) {#3} }
\newcommand{\PLB}[3]{{\it Phys. Lett.} {\bf B{#1}} (19{#2}) {#3}}
\newcommand{\PLBM}[3]{{\it Phys. Lett.} {\bf B{#1}} (20{#2}) {#3}}

\newcommand{\PTP}[3]{{\it Prog. Theor. Phys.} {\bf {#1}} (19{#3}) {#2}}
\newcommand{\PTPM}[3]{{\it Prog. Theor. Phys.} {\bf {#1}} (20{#3}) {#2}}
\newcommand{\ANN}[3]{{\it Ann. Phys. (N.Y.)} {\bf {#1}}, {#2} (19{#3})}

\documentclass[12pt]{article}
\usepackage{amsmath}
\usepackage{amssymb}
\usepackage{graphicx}
\begin{document}
\baselineskip=18pt
\begin{titlepage}
\begin{flushright}
KOBE-TH-13-11\\
\end{flushright}
\vspace{1cm}
\begin{center}{\Large\bf 
The Standard Model \\
with 
\\
New Order Parameters at Finite Temperature 
}
\end{center}
\vspace{1cm}
\begin{center}
Makoto Sakamoto$^{(a)}$
\footnote{E-mail: dragon@kobe-u.ac.jp} and
Kazunori Takenaga$^{(b)}$
\footnote{E-mail: takenaga@kumamoto-hsu.ac.jp}
\end{center}
\vspace{0.2cm}
\begin{center}
${}^{(a)}$ {\it Department of Physics, Kobe University, 
Rokkodai Nada, Kobe, 657-8501 Japan}
\\[0.2cm]
${}^{(b)}$ {\it Faculty of Health Science, Kumamoto
Health Science University, Izumi-machi, Kita-ku, Kumamoto 861-5598, Japan}
\end{center}
\vspace{1cm}
\begin{abstract}
We consider the finite temperature effective potential of the standard model
at the one-loop level in four dimensions by taking account of two kinds of order 
parameters, the Higgs vacuum expectation value and the zero modes
of gauge fields for the Euclidean time direction.
We study the vacuum structure of the model, focusing on 
the existence of new phase, where the zero
modes, that is, the new order parameters develop nontrivial vacuum expectation 
values except for the center of the gauge group. We find that under certain 
conditions there appears no new phase at finite temperature.
\end{abstract}
\end{titlepage}
\section{Introduction}
Quantum field theory at finite temperature \cite{finitet} provides useful tools to
study phase transition in high energy physics. The effective potential at
finite temperature actually plays an important role for studying the 
scenario of the electroweak baryogenesis \cite{baryon} and the deconfinement 
phase of QCD\cite{qcd} in weak coupling regime. Quantum field theory at finite temperature 
has been used in various contexts \cite{example}.

The imaginary time formulation of quantum field theory at finite temperature
is familiar, and in this formulation the Euclidean time direction  
is compactified on a circle $S_{\tau}^1$ whose length of the circumference is
the inverse temperature $T^{-1}$.  When one considers gauge theory on 
such a space, it is well-known that zero modes of component 
gauge fields for the $S_{\tau}^1$ direction cannot be gauged away and 
become dynamical degrees of freedom \cite{qcd}, so that 
they can develop vacuum expectation values \cite{hosotani}. 
We can determine the vacuum expectation values by 
minimizing the effective potential for the zero modes. One should notice 
that such zero modes must be taken into 
account as long as they are the dynamical degrees of freedom.

In the context of higher dimensional gauge theory at finite temperature
zero modes of component gauge fields for the $S_{\tau}$ direction should be taken into
account in addition to possible zero modes of component gauge
fields corresponding to topological spatial extra dimensions. 
Gauge symmetry breaking through the zero modes has been 
discussed in \cite{kobe}\cite{sakatake}. It seems, however, that in the studies of the finite 
temperature phase transition of electroweak models in four 
dimensions, the zero modes of the $SU(2)_L, U(1)_Y$ gauge fields 
for the $S_{\tau}^1$ direction have been overlooked so far though they are dynamical
variables.

One may think that the zero mode of the $SU(2)_L$ gauge fields for the $S_{\tau}$ direction 
takes the value at the center of the $SU(2)_L$ gauge group like QCD 
in weak coupling regime at finite temperature \cite{qcd}. This is, however, not so 
trivial because the models contain the Higgs potential and the 
vacuum expectation value of the Higgs field may influence 
the location of the minimum for the zero mode in the effective potential. This is 
actually the case in physics with extra dimensions \cite{kobe}.

In this paper we investigate the phase structure of the 
standard model in four dimensions at 
finite temperature by studying the effective potential at the one-loop
level. In doing it we correctly take the zero 
modes of the $SU(3)_c, SU(2)_L, U(1)_Y$ gauge fields for 
the $S_{\tau}^1$ direction into account in addition to the usual order 
parameter, the vacuum expectation value of the Higgs field. It is expected 
that there appear new phases, in which the zero modes of the $SU(2)_L, U(1)_Y$ gauge
fields, that is, new order parameters in the model, take 
nontrivial values except for the center of the gauge group. If this is the case, the zero 
modes give a source for the gauge symmetry breaking. We 
focus on seeking whether such a new phase appears or not.

One encounters the situation that has never seen before due to the new order 
parameters. The parametrization of the vacuum expectation value of the Higgs 
field changes, contrary to the usual case. The electromagnetic component in 
the Higgs field, which is usually gauged away by using the $SU(2)_L\times U(1)_Y$ 
degrees of freedom, remains even after using the gauge 
degrees of freedom because of the new order parameters. As a result the number 
of the order parameters increases in the model. We follow the standard prescription to calculate the 
effective potential. We expand fields around the vacuum expectation values and take up to
quadratic terms. The increased order parameters complicate the quadratic 
terms, which contain the couplings that 
break the electromagnetic $U(1)$, denoted by $U(1)_{\rm em}$, invariance.
This makes hard to obtain the effective potential in analytic way.

We impose plausible conditions among the order parameters in order 
to study the effective potential as analytical as possible. 
Even under the conditions we can still investigate a possibility of
new phases in which new order parameters take
nontrivial values. Our analyses tell us  
that there is no new phase in the 
standard model in four dimensions. It may be essential for the result that the boundary 
condition of fermions for the $S_{\tau}^1$ direction is 
fixed by the Fermi statistics. This is 
quite different from the case of the physics with spatial extra dimensions.

This paper is organized as follows. In section $2$ we introduce the 
order parameters of the model and discuss the minimum of the 
tree-level potential for latter convenience. We obtain the
effective potential at the one-loop level under certain conditions among 
the order parameters and study the phase structure by minimizing
the effective potential, focusing on the new phase in section $3$. Conclusions 
and discussions are devoted to section $4$. In Appendix we present the detail of the 
calculations in the presence of the new order parameters.
\section{Order parameters} 
The imaginary time formulation of quantum field theory at finite temperature is to
consider the theory on $S_{\tau}^1\times M^3$, where the Euclidean time direction $\tau$
is compactified on the $S_{\tau}^1$ whose circumference is the inverse temperature $T^{-1}$. 
The $M^3$ is the three-dimensional flat space whose coordinate is denoted by $x^i~(i=1, 2, 3)$.

We consider the standard model in four dimensions at finite temperature. As 
discussed in the literatures \cite{qcd}\cite{hosotani}, the zero modes of the Euclidean time 
components of the gauge fields, which cannot be gauged away, become
the dynamical variable to parametrize the vacuum of the 
theory. They are order parameters of the theory. The 
vacuum expectation values are determined by minimizing 
the effective potential for the order parameters.

In the present case the order parameters we have to take into account are 
\begin{equation}
\vev{A_{\tau}},~~\vev{B_{\tau}},~~\vev{G_{\tau}},~~\vev{\Phi},
\label{shiki1}
\end{equation}
where $A_{\tau}~(B_{\tau}, G_{\tau})$ is the Euclidean time
component of the $SU(2)_L~(U(1)_Y, SU(3)_c)$ gauge field and $\Phi$ is the Higgs field.

Let us discuss the parametrization of the vacuum expectation value (\ref{shiki1}) in 
the electroweak sector. By using 
the $SU(2)_L\times U(1)_Y$ degrees of freedom, we can parametrize the vacuum 
expectation values as
\begin{equation}  
{g\over T}\vev{A_{\tau}}=2\pi~ {\rm diag.}(\varphi, -\varphi),
\quad
{g_Y\over T}\vev{B_{\tau}}=2\pi \theta,
\quad
\vev{\Phi}={1\over\sqrt{2}}
\begin{pmatrix}p \\
v\end{pmatrix}.
\label{shiki2}
\end{equation}
Here $g, g_Y$ are the $SU(2)_L, U(1)_Y$ gauge couplings, respectively.
Let us note that $\varphi$ (also $\theta$) is physically identical to 
$\varphi+l (l\in \mathbb{Z})$. The $\varphi, \theta, v, p$ are real parameters. One can choose another 
parametrization, but equivalent to (\ref{shiki2}), given by
\begin{equation}
{g\over T}\vev{A_{\tau}}=2\pi \begin{pmatrix}
\varphi _3&\varphi_1\\
\varphi_1& -\varphi_3 \\
\end{pmatrix}, \quad 
{g_Y\over T}\vev{B_{\tau}}=2\pi \theta,\quad
 \vev{\Phi}={1\over\sqrt{2}}
 \begin{pmatrix}0 \\
v'\end{pmatrix}.
\label{shiki3}
\end{equation}
The two parametrizations (\ref{shiki2}) and (\ref{shiki3}) are mutually related 
by the transformations
$$
\vev{\Phi'}=V\vev{\Phi},~~~\vev{A_{\tau}'}=V\vev{A_{\tau}}V^{\dagger},
$$
where $V$ is defined by
\begin{equation}
V=
{1\over\sqrt{p^2+v^2}}
\begin{pmatrix}
v & -p\\
p & v
\end{pmatrix}\quad\mbox{with}
\quad VV^{\dagger}=V^{\dagger}V=1,~~{\rm det}~V=1.
\label{shiki4}
\end{equation}
One easily finds that
\begin{equation}
\varphi_1\equiv {2vp\over v^2+p^2},\quad
\varphi_3\equiv {v^2-p^2\over v^2+p^2}\varphi,\quad v'=\sqrt{v^2+p^2}. 
\label{shiki5}
\end{equation}
We employ the parametrization (\ref{shiki2}) in the paper. Let us note that in the vacuum 
expectation value of the Higgs field there remains the
component $p$ that can break the electromagnetic $U(1)$, denoted by $U(1)_{\rm em}$, invariance 
contrary to the usual case where $\vev{A_{\tau}}$ and $\vev{B_{\tau}}$ are not taken into account.

In the $SU(3)_c$ sector, one can parametrize $\vev{G_{\tau}}$ as
\begin{equation}
{g_s\over T}\vev{G_{\tau}}=2\pi~{\rm diag.}(\omega_1, \omega_2, \omega_3)\quad
\mbox{with}\quad \sum_{r=1}^3\omega_r=0.
\label{shiki6}
\end{equation}
$g_s$ is the the $SU(3)_c$ gauge coupling constant. $\omega_r(r=1,2,3)$ is physically identical
to $\omega_r +l (l\in \mathbb{Z})$.

Let us discuss the potential at the tree level. The Higgs potential is given by
\begin{equation}
V_{\rm H}=-\mu^2\Phi^{\dagger}\Phi + {\lambda\over 2}\left(\Phi^{\dagger}\Phi\right)^2.
\label{shiki7}
\end{equation}
In the background of (\ref{shiki2}), the potential at the tree-level is given by 
the Higgs potential (\ref{shiki7}) and the contribution from the Higgs kinetic 
term which yields the third term below,
\begin{eqnarray}
V_{\rm H}^{\rm tree}&=&-\mu^2\abs{\vev{\Phi}}^2+{\lambda\over 2}\left(\abs{\vev{\Phi}}^2\right)^2+
\vev{\Phi}^{\dagger}\left(g\vev{A_{\tau}}+{g_Y\over 2}\vev{B_{\tau}}\right)^2\vev{\Phi}
\nonumber\\
&=&-{\mu^2\over 2}(p^2+v^2)+{\lambda\over 8}(p^2+v^2)^2
+{(2\pi T)^2\over 2}\left\{\left(\varphi +{\theta\over 2}\right)^2p^2
+\left(\varphi -{\theta\over 2}\right)^2v^2\right\}.
\label{shiki8}
\end{eqnarray}
There are three extreme points,
\begin{eqnarray}
{\rm (I)}&:&p=0,~~v=\sqrt{{2\mu^2\over \lambda}},~~\varphi=\theta=0,
\label{shiki9}\\
{\rm (II)}&:&p=v=0,~~\theta={2\mu\over 2\pi T},~~\varphi=0,
\label{shiki10}\\
{\rm (III)}&:&p=v=0,~~\theta=0, ~~\varphi={\mu\over 2\pi T}.
\label{shiki11}
\end{eqnarray}
It is easy to show that (\rm{I}) is the vacuum configuration, and 
(\rm II) and (\rm III) are the saddle point configurations. 
The configuration {\rm (I)} is the usual vacuum in the standard model.

For latter convenience let us minimize the potential under the assumption $p=0$,
\begin{equation}
V_{\rm H}^{\rm tree}\big|_{p=0}=
-{\mu^2\over 2}v^2+{\lambda\over 8}v^4+{(2\pi T)^2\over 2}\left(\varphi - {\theta\over 2}\right)^2v^2.
\label{shiki12}
\end{equation}
There are two extreme points,
\begin{eqnarray}
{\rm (I)}&:&v=\sqrt{{2\mu^2\over \lambda}},~~\varphi-{\theta\over 2}=0,
\label{shiki13}\\
{\rm (II)}&:&
v=0,~~\left(\varphi-{\theta\over 2}\right)^2=\left({\mu\over 2\pi T}\right)^2.
\label{shiki14}
\end{eqnarray}
It is easy to show that the configuration (\rm II) is a saddle point and the vacuum 
configuration is given by (\rm I). Let us 
note that, as long as the second equation in Eq.  (\ref{shiki14}) is
satisfied, arbitrary configurations for $\varphi, \theta$ are allowed.
%
%
\section{One-loop effective potential}
The one-loop effective potential is obtained by the standard prescription. To this end 
one needs to expand the fields around the vacuum expectation values (\ref{shiki2}) and 
takes quadratic terms with respect to fluctuations. The calculation is straightforward, but a little bit 
tedious because of the new order parameters. Namely in the present case, as discussed in the previous 
section,  there remains the component $p$ in the vacuum expectation values of the Higgs field 
that can break the $U_{\rm em}(1)$ invariance and accordingly this results the
couplings that do not conserve the $U(1)_{\rm em}$ charge. This 
never happened in the past calculations of the standard model
at finite temperature. This is entirely due to the new order parameters, that is, the 
vacuum expectation values for $A_{\tau}$. We 
present the details of the calculations in Appendix.

The quadratic terms are given by Eqs. (\ref{shiki64}), (\ref{shiki69}), (\ref{shiki76})  
and (\ref{shiki85}) in Appendix. One needs to find the eigenvalues for the 
matrices, $M^2_{\rm gauge}, M^2_{\rm scalar}, M^2_{\rm ghost}, 
M_{\rm quark}$ and $M_{\rm lepton}$ and  
has to sum up all the Matsubara mode labeled by the integer $n$ whose 
dependence in the eigenvalues is quite nontrivial in the present case. It
may be also difficult to carry out the summation with respect to $n$ 
though we obtain the eigenvalues. The matrices are 
too complex to calculate the effective potential as analytically as 
possible because of the increased order parameters.

We are very much interested in the possibility whether the
new order parameters, namely, $\varphi, \theta$, take the nontrivial values or not.
Taking account of the fact that at the tree level there is no vacuum which breaks 
the $U(1)_{\rm em}$ invariance, it is likely that perturbative corrections
do not induce the vacuum that breaks the $U(1)_{\rm em}$ invariance. Therefore it may 
be natural to assume $p=0$. Since the tree-level potential has the 
global minimum (\ref{shiki14}) under the 
assumption $p=0$ as shown in section $2$, let us impose an ansatz, which is given by
\begin{equation}
p=0,\qquad \varphi-{\theta\over 2}=0.
\label{shiki15}
\end{equation}
This drastically simplifies the quadratic terms with the dependence 
on the new order parameters $\varphi, \theta$. And we are able 
to perform the analytic calculations for obtaining the effective potential at the
one-loop level by the standard prescription. The details 
of the calculations under the ansatz are also given in Appendix.

Let us quote relevant results from Appendix where we give notations and present 
details of calculations.  The quadratic terms for the gauge sector are given by
\begin{eqnarray}
{\cal L}_{\rm gauge}^{{\rm EW} (2)}\big|_{\rm ansatz}
&=&
W_i^- \bar D^{W^{\pm}}\delta_{ij}W_j^{+}
+\half (\bar A_i^3, \bar B_i)
\begin{pmatrix}
\bar D^{A^3}&  {gg_Y\over 4}v^2\\[0.3cm]
{gg_Y\over 4}v^2 &\bar D^B
\end{pmatrix}
\delta_{ij}
\begin{pmatrix}
\bar A_j^3\\
\bar B_j
\end{pmatrix}
\label{shiki16}
\end{eqnarray}
where
\begin{eqnarray}
\bar D^{W^{\pm}}&=&\del_l^2
+\Bigl(\del_{\tau}-i(2\pi T)(2\varphi)\Bigr)^2-{g^2\over 4}v^2,
\nonumber\\
\bar D^{A^3}&=&\del_l^2+\del_{\tau}^2-{g^2\over 4}v^2,\quad 
\bar D^{B}=\del_l^2 +\del_{\tau}^2-{g_Y^2\over 4}v^2.
\label{shiki17}
\end{eqnarray}
As we can see, the second term in Eq. (\ref{shiki16}) can be diagonalized by the usual
rotation defined by Eq. (\ref{shiki96}) in Appendix. The eigenvalues are 
given by Eq. (\ref{shiki97}) in Appendix. The new order parameter $\varphi$ 
appears in the $\bar D^{W^{\pm}}$. Then the one-loop contributions 
from the gauge sector are given by
\begin{eqnarray}
V_{\rm gauge}^{\rm one-loop}&=&
6\times {1\over 2i}\int_k {\rm ln}\biggl[k_i^2 +(2\pi T)^2(n-2\varphi)^2 +{g^2\over 4}v^2\biggr]
\nonumber\\
&+&
3\times {1\over 2i}\int_k {\rm ln}\biggl[k_i^2 +(2\pi T)^2n^2 +{g^2+g_Y^2\over 4}v^2\biggr]
\nonumber\\
&+&
3\times {1\over 2i}\int_k {\rm ln}\biggl[k_i^2 +(2\pi T)^2n^2 \biggr].
\label{shiki18}
\end{eqnarray}
where we have defined 
\begin{equation}
\int_k \equiv iT \sum_{n=-\infty}^{\infty}\int {d^3 k\over(2\pi)^3}.
\label{shiki19}
\end{equation}
The integer $n$ stands for the Matsubara mode.

The quadratic terms from the scalar sector under the ansatz, including 
the Euclidean time components of the gauge fields, $\bar A_{\tau}^a, \bar B_{\tau}$ 
are given by
\begin{eqnarray}
{\cal L}_{\rm scalar}^{(2)}\big|_{\rm ansatz}
&=&
\half (\bar A_{\tau}^1, \bar A_{\tau}^2)
\begin{pmatrix}
\bar A& a  \\
-a &\bar A\\
\end{pmatrix}
\begin{pmatrix}
\bar A_{\tau}^1\\
\bar A_{\tau}^2
\end{pmatrix}
+
\half  (g^1, g^2)
\begin{pmatrix}
\bar B& \bar g  \\
-\bar g &\bar C\\
\end{pmatrix}
\begin{pmatrix}
g^1\\
g^2
\end{pmatrix}\nonumber\\
&+&
\half  (\bar A_{\tau}^3, \bar B_{\tau})
\begin{pmatrix}
\bar D& \bar l  \\
\bar l &\bar E\\
\end{pmatrix}
\begin{pmatrix}
\bar A_{\tau}^3\\
\bar B_{\tau}
\end{pmatrix}
+
\half h \bar F h +\half G^0 \bar G G^0.
\label{shiki20}
\end{eqnarray}
where
\begin{eqnarray}
\bar A&=&\del_i^2 +\del_{\tau}^2 -(2\pi T)^2 (2\varphi)^2-{g^2\over 4}v^2,
\nonumber\\
\bar B&=&\del_i^2 +\del_{\tau}^2 -(2\pi T)^2 \left(2\varphi\right)^2+\mu^2
-{\lambda\over 2}v^2 -{g^2\over 4}v^2 ~~=\bar C, 
\nonumber\\
%
%
\bar D&=&\del_i^2 +\del_{\tau}^2-{g^2\over 4}v^2, ~~
\bar E=\del_i^2 +\del_{\tau}^2-{g_Y^2\over 4}v^2, 
\nonumber\\
\bar F&=&\del_i^2 +\del_{\tau}^2+\mu^2 - {3\over 2}\lambda v^2, 
\nonumber\\
\bar G&=&\del_i^2 +\del_{\tau}^2 +\mu^2-{\lambda v^2\over 2}-{g^2+g_Y^2\over 4}v^2,
\nonumber\\
a&=&-2(2\pi T)(2\varphi)\del_{\tau},\quad
\bar g=-2(2\pi T)\left(2\varphi\right)\del_{\tau}, \quad \bar l={gg_Y\over 4}v^2 .
\label{shiki21}
\end{eqnarray}
The first and the second terms in Eq. (\ref{shiki20}) is automatically diagonalized by the original
complex base,
\begin{equation}
W_{\tau}^{\pm}={1\over\sqrt{2}}(\bar A_{\tau}^1\mp i \bar A_{\tau}^2),\quad
G^{\pm}={1\over\sqrt{2}}(g^1\mp ig^2).
\label{shiki22}
\end{equation} 
The third term is diagonalized by the usual rotation by Eq. (\ref{shiki96}) as 
before. The eigenvalues are given by Eq. (\ref{shiki102}). Then the contributions 
from the scalar sector are
\begin{eqnarray}
V_{\rm scalar}^{\rm one-loop}&=&
2\times {1\over 2i}\int_k {\rm ln}\biggl[k_i^2 +(2\pi T)^2(n-2\varphi)^2 +{g^2\over 4}v^2\biggr]
\nonumber\\
&+&
1\times {1\over 2i}\int_k {\rm ln}\biggl[k_i^2 +(2\pi T)^2n^2 +{g^2+g_Y^2\over 4}v^2\biggr]
\nonumber\\
&+&
1\times {1\over 2i}\int_k {\rm ln}\biggl[k_i^2 +(2\pi T)^2n^2 \biggr]
\nonumber\\
&+&
2\times {1\over 2i}\int_k {\rm ln}\biggl[k_i^2 +(2\pi T)^2(n-2\varphi)^2-\mu^2 
+{\lambda\over 2}v^2 +{g^2\over 4}v^2\biggr]
\nonumber\\
&+&
1\times {1\over 2i}\int_k {\rm ln}\biggl[k_i^2 +(2\pi T)^2n^2 -\mu^2
+{\lambda\over 2}v^2+{g^2+g_Y^2\over 4}v^2\biggr]
\nonumber\\
&+&
1\times {1\over 2i}\int_k {\rm ln}\biggl[k_i^2 +(2\pi T)^2n^2 -\mu^2
+{3\lambda\over 2}v^2\biggr].
\label{shiki23}
\end{eqnarray}

The quadratic terms from the ghost sector are given by
\begin{eqnarray}
{\cal L}_{\rm ghost}^{(2)}\big|_{\rm ansatz}
&=&
-i\bar C^+ \bar D^{W^{\pm}} C^-
-i\bar C^- \bar D^{W^{\pm}} C^+
-i(\bar C^3, \bar C)
\begin{pmatrix}
\bar D^{A^3}& {gg_Y\over 4}v^2 \\[0.3cm]
{gg_Y\over 4}v^2 & \bar D^B\\
\end{pmatrix}
\begin{pmatrix}
C^3\\
C 
\end{pmatrix},
\label{shiki24}
\end{eqnarray}
where $\bar D^{W^{\pm}}, \bar D^{A^3}, \bar D^B$ are the same as the 
ones given in Eq. (\ref{shiki17}). Then the contributions to the effective potential are
\begin{eqnarray}
V_{\rm ghost}^{\rm one-loop}&=&
-4\times {1\over 2i}\int_k {\rm ln}\biggl[k_i^2 +(2\pi T)^2(n-2\varphi)^2 +{g^2\over 4}v^2\biggr]
\nonumber\\
&-&
2\times {1\over 2i}\int_k {\rm ln}\biggl[k_i^2 +(2\pi T)^2n^2 +{g^2+g_Y^2\over 4}v^2\biggr]
\nonumber\\
&-&
2\times {1\over 2i}\int_k {\rm ln}\biggl[k_i^2 +(2\pi T)^2n^2 \biggr].
\label{shiki25}
\end{eqnarray}
We observe from Eqs. (\ref{shiki18}) and (\ref{shiki25}) that the 
on-shell degrees of freedom for the gauge fields are extracted by the ghost fields. 
Let us note that the ghost fields obey the periodic boundary conditions \cite{hatakugo}.

The $SU(3)_c$ gauge contribution to the effective potential is given from 
Eq. (\ref{shiki92}) in Appendix by
\begin{equation}
V_{SU(3)_c}^{\rm one-loop}=
(4-2)\sum_{r, q=1}^3{1\over 2i}\int_k{\rm ln}\biggl[k_i^2+(2\pi T)^2(n+\omega_r -\omega_q)^2\biggr].
\label{shiki26}
\end{equation}

We consider only the third generation for fermions. Our results do not change even if
we introduce the first and second generations with the mixings among the generations.  
And we simply assume that the neutrino is massless. Then the quadratic terms from the fermions 
are given by
\begin{eqnarray}
{\cal L}_{\rm fermion}^{(2)}\big|_{\rm ansatz}&=&
(\bar t_L, \bar t_R)
\begin{pmatrix}
-i\bar D_{t_L}& {f_t\over\sqrt{2}}v \\
{f_t\over\sqrt{2}}v  & -i\bar D_{t_R}\\
\end{pmatrix}
\begin{pmatrix}
t_L\\
t_R
\end{pmatrix}
+
(\bar b_L, \bar b_R)
\begin{pmatrix}
-i\bar D_{b_L}& {f_b\over\sqrt{2}}v \\
{f_b\over\sqrt{2}}v  & -i\bar D_{b_R}\\
\end{pmatrix}
\begin{pmatrix}
b_L\\
b_R
\end{pmatrix}
\nonumber\\
&+&
(\bar \tau_L, \bar \tau_R)
\begin{pmatrix}
-i\bar D_{{\tau}_L}& {f_{\tau}\over\sqrt{2}}v \\
{f_{\tau}\over\sqrt{2}}v  & -i\bar D_{{\tau}_R}\\
\end{pmatrix}
\begin{pmatrix}
\tau_L\\
\tau_R
\end{pmatrix}
-i\bar \nu_L \bar D_{\nu_L}\nu_L ,
\label{shiki27}
\end{eqnarray}
where
\begin{eqnarray}
\bar D_{t_L}&\equiv &\gamma_{\tau}\Bigl(\del_{\tau}-i(2\pi T)\bigl(\omega_r +
{4\over 3}\varphi\bigr)\Bigr)+\gamma_i\del_i~~
=\bar D_{t_R},\nonumber\\
\bar D_{b_L}&\equiv &\gamma_{\tau}\Bigl(\del_{\tau}-i(2\pi T)\bigl(\omega_r-{2\over 3}\varphi \bigr)\Bigr)
+\gamma_i\del_i ~~=\bar D_{b_R},\nonumber\\
\bar D_{\tau_L}&\equiv &\gamma_{\tau}\Bigl(\del_{\tau}+i(2\pi T)2\varphi \Bigr)+\gamma_i\del_i
~~=\bar D_{\tau_R},
\nonumber\\
\bar D_{\nu_L}&\equiv &\gamma_{\tau}\del_{\tau}+\gamma_i\del_i.
\label{shiki28}
\end{eqnarray}
The contributions to the effective potential are given by the determinants of the two 
by two matrices in Eq. (\ref{shiki27}) and by taking the logarithm of them, which is given by
\begin{eqnarray}
V_{\rm fermion}^{\rm one-loop}&=&
(-1)2^2\times {1\over 2i}\sum_{r=1}^3\int_k {\rm ln}\biggl[k_i^2 +(2\pi T)^2
\bigl(n+\half -\omega_r -{4\over 3}\varphi\bigr)^2 +{f_t^2\over 2}v^2\biggr]
\nonumber\\
&+&
(-1)2^2\times {1\over 2i}\sum_{r=1}^3\int_k {\rm ln}\biggl[k_i^2 +(2\pi T)^2
\bigl(n+\half -\omega_r +{2\over 3}\varphi\bigr)^2  +{f_b^2\over 2}v^2\biggr]
\nonumber\\
&+&
(-1)2^2 \times {1\over 2i}\int_k {\rm ln}\biggl[k_i^2 +(2\pi T)^2
\bigl(n+\half +2\varphi\bigr)^2  +{f_{\tau}^2\over 2}v^2\biggr]
\nonumber\\
&+&
(-1){2^2\over 2}\times {1\over 2i}\int_k {\rm ln}\biggl[k_i^2 +(2\pi T)^2\bigl(n+\half\bigr)^2 \biggr],
\label{shiki29}
\end{eqnarray}
Let us note that the half-integer in the Matsubara mode $n$ comes from the antiperiodic
boundary condition of fermion for the $S_{\tau}^1$ direction. Collecting 
all the contributions obtained above, we obtain the effective potential at the 
one-loop level under the ansatz (\ref{shiki15}), as given by Eq. (\ref{shiki111}) in Appendix.

The typical expression we have to evaluate is 
\begin{equation}
V_{\rm basics}=(-1)^FN_{\rm deg.}\left({1\over 2i}\right)
\int_k \ln \left(k^2+(2\pi T)^2(n+\varphi)^2+m(v)^2\right),
\label{shiki30}
%
%
\end{equation}
where $F$ takes $1(0)$ for fermions (bosons) and $N_{\rm deg.}$ counts the degrees of freedom.
Following the standard prescription, $V_{\rm basics}$ consists of 
the zero temperature part and the finite temperature part,  
\begin{equation}
V_{\rm basics}=V_{\rm basics}^{T=0} + V_{\rm basics}^{T\neq 0},
\label{shiki31}
\end{equation}
where
\begin{eqnarray}
V_{\rm basics}^{T=0}&=&
-(-1)^{F+1}N_{\rm deg} {m(v)^4\over 4(4\pi)^2}\left(\ln \left({m(v)^2\over M^2}\right)-{3\over 2}\right),
\label{shiki32}\\
V_{\rm basics}^{T\neq 0}
&=&(-1)^{F+1}N_{\rm deg}{2\over (2\pi)^2}T^4\sum_{m=1}^{\infty}
{1\over m^4}\cos\left(2\pi m\varphi\right)
\left({m(v)^2\over T^2}m^2\right)K_2\left({m(v)\over T}m\right).
\label{shiki33}
\end{eqnarray}
Let us note that the Matsubara mode $n$ is now replaced by the \lq\lq winding\rq\rq~mode $m$ through
the Poisson's resummation formula,
\begin{equation}
\sqrt{4\pi t}~T\sum_{n=-\infty}^{\infty}\e^{-t(2\pi T)^2(n+\varphi)^2}
=\sum_{m=-\infty}^{\infty}\e^{-\bigl({m^2\over 4T^2}{1\over t}+2\pi i m \varphi\bigr)}.
\label{Shiki34}
\end{equation}
Here we have employed the $\overline{MS}$ scheme
for the zero temperature part of the effective potential, which comes from the $m=0$ 
mode and $M$ is a certain mass scale. $K_2(z)$ is the modified Bessel function defined by
\begin{equation}
\int_0^{\infty}dt~t^{-\nu -1}\e^{-At-{B\over t}}=2\left({A\over B}\right)^{\nu\over 2}K_2(2\sqrt{AB}).
\label{shiki35}
\end{equation}
Equipped with Eqs. (\ref{shiki32}) and (\ref{shiki33}), we finally obtain the
effective potential at the one-loop level, as given by 
Eqs. (\ref{shiki115}) and (\ref{shiki116}) in Appendix. Let us 
note that the finite temperature part (\ref{shiki33}) becomes the 
same as the one obtained by Dolan and 
Jackiw \cite{finitet} for $\varphi=0$ and $m(v)/T\ll 1$ by expanding the
modified Bessel function in polynomial \cite{niemi}.

We are very much interested in the new phase, in which the new order 
parameter $\varphi$ takes the nontrivial value except for the center of gauge group. 
The new order parameters $\varphi$ and $\omega_r$ appear only 
in the finite temperature parts of the contributions 
from the $W^{\pm}, G^{\pm}, t, b, \tau, G_{\mu}$, which is 
given from Eq. (\ref{shiki114}) by
\begin{eqnarray}
&&V^{T\neq 0}_{\varphi,~\omega_r -{\rm dep.}}\nonumber\\
&=&-4{2\over (2\pi)^2}T^4\sum_{m=1}^{\infty}
{1\over m^4}\cos[2\pi m(2\varphi)]\left({m_W(v)^2\over T^2}m^2\right)
K_2\left({m_W(v)\over T}m\right)
\nonumber\\
&-&2{2\over (2\pi)^2}T^4\sum_{m=1}^{\infty}
{1\over m^4}\cos[2\pi m(2\varphi)]
\left({m_{G^{\pm}}(v)^2\over T^2}m^2\right)K_2\left({m_{G^{\pm}}(v)\over T}m\right)
\nonumber\\
&+&4{2\over (2\pi)^2}T^4\sum_{r=1}^3\sum_{m=1}^{\infty}
{(-1)^m\over m^4}\cos\Bigl[2\pi m\bigr(\omega_r + {4\over 3}\varphi\bigl)\Bigr]
\left({m_{t}(v)^2\over T^2}m^2\right)K_2\left({m_{t}(v)\over T}m\right)
\nonumber\\
&+&4{2\over (2\pi)^2}T^4\sum_{r=1}^3\sum_{m=1}^{\infty}
{(-1)^m\over m^4}\cos\Bigl[2\pi m\bigr(\omega_r -{2\over 3}\varphi\bigl)\Bigr]
\left({m_{b}(v)^2\over T^2}m^2\right)K_2\left({m_{b}(v)\over T}m\right)
\nonumber\\
&+&4{2\over (2\pi)^2}T^4\sum_{m=1}^{\infty}
{(-1)^m\over m^4}\cos[2\pi m(2\varphi)]
\left({m_{\tau}(v)^2\over T^2}m^2\right)K_2\left({m_{\tau}(v)\over T}m\right)
\nonumber\\
&-&2{2\over(2\pi)^2}T^4\sum_{r, q=1}^3
\sum_{m=1}^{\infty}{2\over m^4}\cos[2\pi m(\omega_r - \omega_q)].
\label{shiki36}
\end{eqnarray}
The notation $m_i(v)~(i=W, G^{\pm}, t, b, \tau)$ is defined by Eq. (\ref{shiki117}) in Appendix.

Let us minimize $V^{T\neq 0}_{\varphi,~\omega_r -{\rm dep.}}$ with respect to $\varphi, \omega_r$.
It has been well-known that the $SU(3)_c$ gauge sector, the last 
line in Eq. (\ref{shiki36}), is minimized at
\begin{equation}
\omega_r= {k\over 3}\quad (k=0,1,2)\quad (\mbox{mod}~1).
\label{shiki37}
\end{equation}
One sees from the Polyakov loop defined by
\begin{equation}
W_p^{SU(3)_c}={\cal P}{\rm exp}\biggl(ig_s \int_0^{1\over T}~d\tau~\vev{G_{\tau}}\biggr)
\bigg|_{\omega_r={k\over 3}}
=\e^{i2\pi {k\over 3}}~{\bf 1}_{3\times 3}
\label{shiki38}
\end{equation}
that it is the center of $SU(3)_c$ gauge group.

The typical structure of the potential (\ref{shiki36}) 
is given by the following two types of the functions:
\begin{eqnarray}
f(x,\tilde z)&=&-\sum_{m=1}^{\infty}
{1\over m^4}\cos[2\pi m x]
\left({\tilde z}m\right)^2K_2\left(\tilde z m\right),\nonumber\\
g(x, \tilde z)&=&+\sum_{m=1}^{\infty}
{(-1)^m\over m^4}\cos[2\pi m x]
\left({\tilde z}m\right)^2K_2\left(\tilde z m\right).
\label{shiki39}
\end{eqnarray}
We numerically depict $f(x, \tilde z)$ and $g(x,\tilde z)$ for positive 
values of $\tilde z$ in Figs. $1$ and $2$. 
\begin{figure}[htbp]
\begin{minipage}{0.5\hsize}
\begin{center}
\includegraphics[width=65mm]{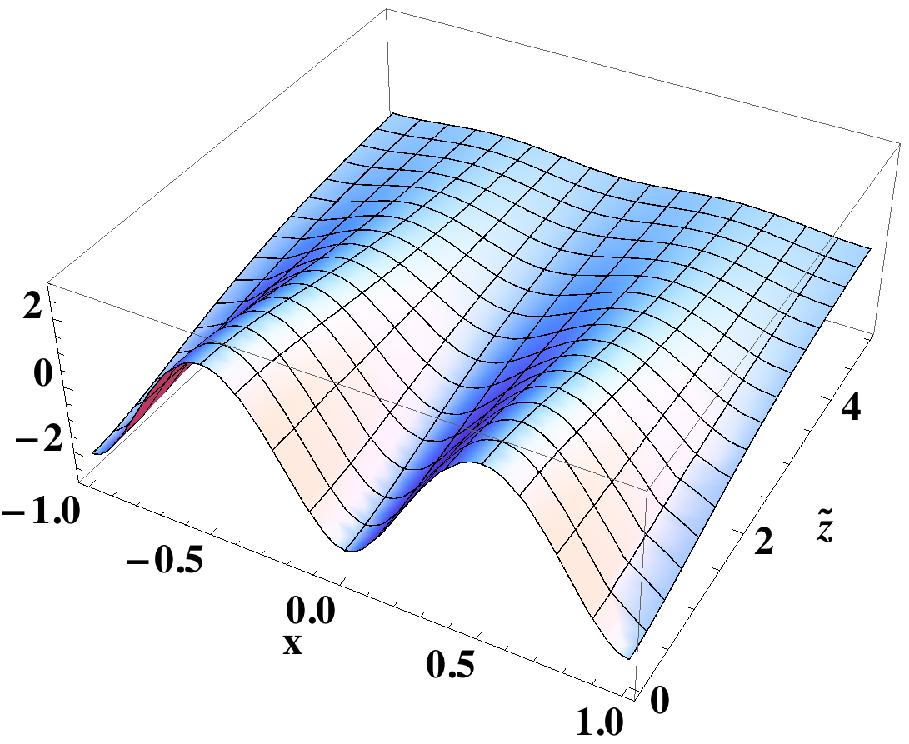}
\end{center}
\caption{The behavior of $f(x,\tilde z)$.}
\end{minipage}
\begin{minipage}{0.5\hsize}
\begin{center}
\includegraphics[width=65mm]{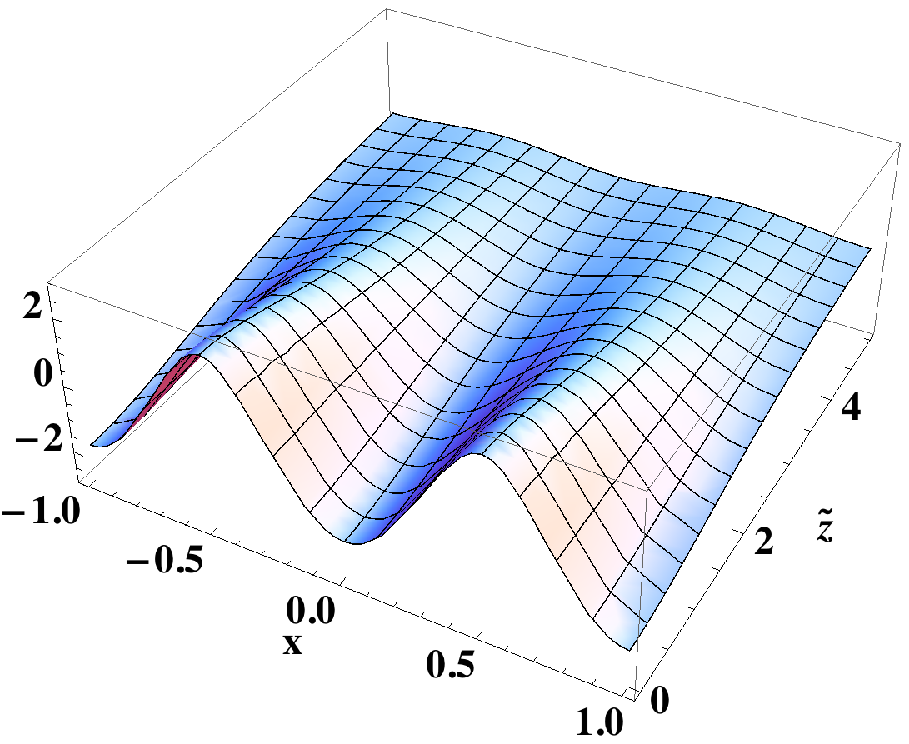}
\end{center}
\caption{The behavior of $g(x,\tilde z)$.}
\end{minipage}
\end{figure}
We find that both of the functions are
minimized at $x=0$ for $\tilde z >0$. This result will be understood by noting that 
the $m=1$ mode dominates the functions $f(x,\tilde z)$ and $g(x,\tilde z)$ to 
yield $x=0$ as the minimum configuration.
%
%
It is also confirmed numerically that the $m=1$ mode dominates the two functions. 
Then the configurations that minimize  
$V^{T\neq 0}_{\varphi,~\omega_r-{\rm dep.}}$ are given by
\begin{eqnarray}
&&\omega_r+{4\over 3}\varphi=0~(r=1,2,3),~~\omega_r-{2\over 3}\varphi=0~(r=1,2,3)
,\quad 2\varphi=0~~(\mbox{mod}~1)
\label{shiki41}
\end{eqnarray}
and (\ref{shiki37}). The first and the second equations in Eq. (\ref{shiki41}) are resulted by 
the top, bottom contributions and the third one is resulted 
by $W^{\pm}, G^{\pm}$ and tau contributions. We immediately 
find that $V^{T\neq 0}_{\varphi,~\omega_r -{\rm dep.}}$ is minimized 
at $\varphi =\tilde k/2 (\tilde k=0,1)$ and $\omega_r=k/3 (k=0,1,2)$ (mod $1$) for nonzero 
temperature. This also implies $\theta=0$ (mod $1$) at the vacuum from the second equation in the ansatz (\ref{shiki15}). 
We conclude that the center 
of $SU(3)_c$ and $SU(2)_L$ is the vacuum configuration at finite temperature .
%
%
%
%

There is no new phase, in which the new order parameters take nontrivial
values other than the center of the gauge group. It 
is crucial that the boundary condition of fermions for the Euclidean time direction
must be antiperiodic due to the Fermi statistics. This is essential 
for $\varphi$ to take the center of $SU(2)_L$ at the minimum of the effective potential. This 
is a remarkable difference when we consider the boundary conditions of the fields for 
extra dimensions, which is {\it a priori} unknown. 	
\section{Conclusions and Discussions}
We have taken account of the new order parameters arising from the
zero mode of the Euclidean time components of the gauge fields for 
studying the effective potential of the standard model 
at finite temperature in four dimensions. Because of the increased number of the order 
parameters, there remains the component that can break the 
electromagnetic $U(1)$, denoted by $U(1)_{\rm em}$, invariance in the 
parametrization of the vacuum expectation values of the Higgs field.

The existence of such a parameter complicates the quadratic terms 
of the fluctuating fields, by which one obtains the effective
potential at the one-loop level. We have imposed the ansatz, which
preserves the $U(1)_{\rm em}$ invariance, in order 
to study the effective potential as analytically as possible. Then we have obtained 
the analytic expression for the effective potential and study the 
vacuum structure, namely, the possibility whether the new
order parameters take nontrivial values except for the center of the gauge 
group or not.

We find that the new order parameters do not take the nontrivial values. 
It is important that the fermion must obey the antiperiodic  
boundary condition for the Euclidean time direction due to the Fermi statistics.
Thanks to this fact, the new order parameter $\varphi$ always takes zero 
at the vacuum for finite temperature. It 
has been pointed out in \cite{kobe} that the nontrivial phase exists if there is a cross term 
between $v$ and $\varphi$ in the tree-level potential.
%
%
One may think that the absence of the cross term in Eq. (\ref{shiki8}) due to the ansatz is 
the reason to exist 
no new phases. Such the cross term, however, is not important for existence of the
new phase. The boundary condition for the compactified direction 
is essential. In the case of \cite{kobe}, the boundary condition can be taken
to be periodic even for fermions because of the spatial compactified 
direction, which, contrary to the Euclidean time direction, is free from 
the Fermi statistics. This makes it possible that the new order parameter $\varphi$ 
can take a nontrivial value except for the center of the gauge group.

Some comments are in order. Originally there are four order 
parameters, $v, p, \varphi, \theta$. Even though it may be unlikely
for $p, \varphi, \theta$ to have nontrivial values at the vacuum, 
one should calculate the effective potential at the one-loop level by using the 
matrices, Eqs. (\ref{shiki66}), (\ref{shiki73}), (\ref{shiki78}), (\ref{shiki86}) 
and (\ref{shiki88}) without imposing any conditions among the order 
parameters. To this end one needs to develop the technique to sum up all 
the Matsubara modes $n$ which have the complicated dependence 
in the eigenvalues of the matrices. Concerning with
this point, we would like to mention the limit $g, g_Y\rightarrow 0$. Even in the
limit the dependence of the quadratic terms on $\varphi, \theta$ is survived in 
the simplified form. Nevertheless it is still difficult to perform the analytic calculation 
for obtaining the effective potential at the one-loop level.

It may be true that the order parameters
arising from the Euclidean components of the gauge fields do not 
develop nontrivial values except for the center of the gauge group. If one considers
two Higgs doublets models, including the minimal supersymmetric standard 
model, for example, the number 
of order parameters in the models is larger than the usual case because of 
the additional zero modes of the Euclidean components of the gauge fields.
Then there may be a possibility of a new source for CP violation at finite
temperature in dependence on the structure of the Higgs 
potential\footnote{Spontaneous CP violation at finite 
temperature has been reported in \cite{funakubo}.}.

%
%
%
%
%
%
%

We stress that as long as the zero mode of the Euclidean time component of 
the gauge field becomes the dynamical variable at finite 
temperature field theory, it is important and
natural to take it into account to study the effective potential at finite 
temperature in addition to the usual  order parameter such as the Higgs field. 
We are now studying the electroweak models such as the two Higgs doublet 
models, including the minimal supersymmetric standard model, at finite temperature 
by taking account of all the order parameters \cite{sakatake2}. We are interested in 
whether the order parameters take the nontrivial 
values or not. This will be reported elsewhere in the future.
%
%
\begin{center}
{\bf Acknowledgement}
\end{center}
This work is supported in part by a Grant-in-Aid for Scientific Research
(No. 24540291 (K.T.)) from the Japanese Ministry of Education, Science, Sports and Culture. 
\vspace*{1cm}
\begin{center}
{\bf \Large Appendix}  
\end{center}
In this Appendix we present the details of notations and calculations, some of which are used in the text.
\begin{flushleft}
{\bf Quadratic terms}
\end{flushleft}
In the imaginary time formulation of finite temperature field theory, the Euclidean time $\tau$
is defined by the Wick rotation
\begin{equation}
\tau = -i t.
\label{shiki42}
\end{equation}
Here we use $\eta^{\mu\nu}={\rm diag.}(1,-1,-1,-1)$ as the Minkowski metric.
Accordingly, the Euclidean time component of the gauge field is related with the Minkowski 
component of the gauge field by
\begin{equation}
A_{\tau}=iA_0,\qquad B_{\tau}=iB_0, \qquad G_{\tau}= iG_0.
\label{shiki43}
\end{equation}

The Euclidean time direction is compactified on $S_{\tau}^1$ whose circumference is
the inverse temperature $T^{-1}$. Bosons (Fermions) must obey 
the (anti) periodic boundary conditions because of quantum statistics. Then the Euclidean
component of the momentum $k_{\tau}=-ik_0$ is 
discretized as
\begin{equation}
k_{\tau}=\left\{
\begin{array}{ll}
\omega_n^B=2\pi T n &\mbox{for~bosons},\\
\omega_n^F=2\pi T\left(n+\half\right)&\mbox{for~fermions},
\label{shiki44}
\end{array}
\right.
\end{equation}
where $n$ denotes the Matsubara mode, $n=0,\pm 1, \pm 2, \cdots$.

In order to obtain the effective potential at the one-loop level,  we expand 
the fields around the vacuum expectation values as defined by 
Eq. (\ref{shiki2}) in the text and take up to the 
quadratic terms with respect to the fluctuations.

Let us start with the gauge sector of the standard model,
\begin{equation}
{\cal L}_{\rm gauge}^{\rm EW}=-\frac{1}{2}{\rm tr}~(F_{\mu\nu}F^{\mu\nu})-
\frac{1}{4}B_{\mu\nu}B^{\mu\nu},
\label{shiki45}
\end{equation}
where $F_{\mu\nu}, B_{\mu\nu}$ are the $SU(2)_L, U(1)_Y$ field strengths, respectively.
We consider the $SU(3)_c$ gauge sector later. The quadratic terms from the gauge kinetic terms 
are obtained by 
\begin{eqnarray}
{\cal L}_{\rm gauge}^{{\rm EW} (2)}&=&
-{1\over 2}\biggl(\del_i\bar A_j^a\del_i\bar A_j^a - \del_i\bar A_j^a \del_j\bar A_i^a\biggr)
-{1\over 2}(D_{\tau}^{SU(2)}\bar A_i)^a(D_{\tau}^{SU(2)}\bar A_i)^a\nonumber\\
&-&{1\over 2}\del_i\bar A_{\tau}^a\del_i\bar A_{\tau}^a
-{1\over 2}\biggl(\del_i \bar B_j\del_i \bar B_j - \del_j\bar B_i \del_i \bar B_j\biggr)
-{1\over 2}\del_i\bar B_{\tau}\del_i\bar B_{\tau}
-{1\over 2}\del_{\tau}\bar B_i\del_{\tau}\bar B_i
\nonumber\\
&+&(D_{\tau}^{SU(2)}\bar A_i )^a\del_i\bar A_{\tau}^a +\del_i\bar B_{\tau}\del_{\tau}\bar B_i~,
\label{shiki46}
\end{eqnarray}
where we have defined 
\begin{equation}
D_{\tau}^{SU(2)}\bar A_i\equiv \del_{\tau}\bar A_i - ig[\vev{A_{\tau}},~\bar A_i]\quad (i=1, 2, 3).
\label{shiki47}
\end{equation}
$a(=1,2,3)$ is the $SU(2)_L$ index and the $i, j, l$ stand for the space component. 
The bar on the field denotes the fluctuation.
The Higgs kinetic term and potential are
\begin{equation}
{\cal L}_{\rm Higgs}=(D_{\mu}\Phi)^{\dagger}D^{\mu}\Phi  - V_{\rm H},
\label{shiki48}
\end{equation}
where
\begin{equation}
D_{\mu}\Phi=\del_{\mu}\Phi - igA_{\mu}\Phi - i{g_Y\over 2}B_{\mu}\Phi,~~
V_{\rm H}=-\mu^2\Phi^{\dagger}\Phi
+{\lambda\over 2}(\Phi^{\dagger}\Phi)^2.
\label{shiki49}
\end{equation}
The quadratic terms of the Higgs sector are given by
\begin{eqnarray}
{\cal L}_{\rm Higgs}^{(2)}&=&
-(\bar D_{\tau}\bar\Phi)^{\dagger}\bar D_{\tau}{\bar\Phi}
-(\del_i \bar\Phi)^{\dagger}\del_i \bar\Phi
\nonumber\\
&-&(\bar D_{\tau}\bar\Phi)^{\dagger}\biggl(-ig\bar A_{\tau}\vev{\Phi}-
i{g_Y\over 2}\bar B_{\tau}\vev{\Phi}\biggr)
-\biggl(ig\vev{\Phi}^{\dagger}\bar A_{\tau}+i{g_Y\over 2}\vev{\Phi}^{\dagger}
\bar B_{\tau}\biggr)\bar D_{\tau}\bar \Phi
\nonumber\\
&-&\del_i \bar\Phi^{\dagger}\biggl(-ig\bar A_i\vev{\Phi}-i{g_Y\over 2}\bar B_i\vev{\Phi}\biggr)
-\biggl(ig\vev{\Phi}^{\dagger}\bar A_i+i{g_Y\over 2}\vev{\Phi}^{\dagger}\bar B_i\biggr)\del_i \bar \Phi
\nonumber\\
&-&g^2\biggl(\vev{\Phi}^{\dagger}\vev{A_{\tau}}\bar A_{\tau}\bar\Phi+
\bar\Phi^{\dagger}\bar A_{\tau}\vev{A_{\tau}}\vev{\Phi}+\vev{\Phi}^{\dagger}\bar A_{\tau}
\bar A_{\tau}\vev{\Phi}\biggr)
\nonumber\\
&-&{gg_Y\over 2}\biggl(\vev{\Phi}^{\dagger}\vev{A_{\tau}}\bar B_{\tau}\bar\Phi + 
\bar\Phi^{\dagger}\bar B_{\tau}\vev{A_{\tau}}\vev{\Phi} + \vev{\Phi}^{\dagger}\bar A_{\tau}
\bar B_{\tau}\vev{\Phi}\biggr)
\nonumber\\
&-&{gg_Y\over 2}\biggl( \vev{\Phi}^{\dagger}\vev{B_{\tau}}\bar A_{\tau}\bar \Phi +
\bar\Phi^{\dagger}\bar A_{\tau}
\vev{B_{\tau}}\vev{\Phi} + \vev{\Phi}^{\dagger}\bar B_{\tau}\bar A_{\tau}\vev{\Phi}\biggr)
\nonumber\\
&-&\left({g_Y\over 2}\right)^2\biggl(\vev{\Phi}^{\dagger}\vev{B_{\tau}}\bar B_{\tau}\bar\Phi +
\bar\Phi^{\dagger}\bar B_{\tau}\vev{B_{\tau}}\vev{\Phi}+\vev{\Phi}^{\dagger}
\bar B_{\tau}\bar B_{\tau}\vev{\Phi}\biggr)
\nonumber\\
&-&g^2\vev{\Phi}^{\dagger}\bar A_i\bar A_i \vev{\Phi}
-gg_Y\vev{\Phi}^{\dagger}\bar B_i \bar A_i\vev{\Phi}
-\left({g_Y\over 2}\right)^2\vev{\Phi}^{\dagger}\bar B_i\bar B_i\vev{\Phi}
\nonumber\\
&-&\mu^2\bar\Phi^{\dagger}\bar\Phi
+{\lambda\over 2}\left(2\abs{\vev{\Phi}}^2\abs{\bar\Phi}^2+
2(\vev{\Phi}^{\dagger}\bar \Phi)(\bar\Phi^{\dagger}\vev{\Phi})
+(\vev{\Phi}^{\dagger}\bar\Phi)^2
+(\bar\Phi^{\dagger}\vev{\Phi})^2\right).
\label{shiki50}
\end{eqnarray}
Here we have defined 
\begin{equation}
\bar D_{\tau}\bar\Phi\equiv \del_{\tau}\bar \Phi - 
ig \vev{A_{\tau}}\bar\Phi - i{g_Y\over 2}\vev{B_{\tau}}\bar\Phi .
\label{shiki51}
\end{equation}

Now let us introduce the gauge fixing and the ghosts,
\begin{equation}
{\cal L}_{{\rm gf} + {\rm FP}}={\cal L}_{{\rm gf}+{\rm FP}}^{SU(2)_L}
+{\cal L}_{{\rm gf}+{\rm FP}}^{U(1)_Y}
=(-i){\delta}_B(\bar C^a F^a)+(-i)\delta_B(\bar CF),
\label{shiki52}
\end{equation}
where $\delta_B$ denotes the 
BRS transformations\cite{kugoojima}. Let us note that $\bar C^a$ and $\bar C$ are 
anti-ghost fields. The gauge fixing functions are chosen to be
\begin{eqnarray}
F^a&\equiv &-\del_i \bar A_i^a-\alpha_1
\biggl[(D_{\tau}^{SU(2)}\bar A_{\tau})^a-
ig \left(\bar\Phi^{\dagger}{\tau^a\over 2}\vev{\Phi}-
\vev{\Phi}^{\dagger}{\tau^a\over 2}\bar\Phi\right)\biggr]+{\alpha_1\over 2}b^a,
\label{shiki53}\\
F&\equiv &-\del_i \bar B_i-\alpha_2\biggl[\del_{\tau}\bar B_{\tau} - i {g_Y\over 2}
\left(\bar\Phi^{\dagger}\vev{\Phi}-\vev{\Phi}^{\dagger}\bar\Phi\right)\biggr]
+{\alpha_2\over 2}b,
\label{shiki54}
\end{eqnarray}
where $\alpha_1$ and $\alpha_2$ are the gauge parameters. Hereafter we 
take $\alpha_1=\alpha_2\equiv \xi$ for simplicity. After operating the BRS 
transformations and performing the integration 
of $b^a$ and $b$ fields, the quadratic terms from the $SU(2)_L$ part are given by
\begin{eqnarray}
&&{\cal L}_{\rm gf}^{SU(2)_L~(2)}=(-i)\delta_B(C^aF^a)\big|_{\rm quadratic }\nonumber\\
&= &-{1\over 2\xi}\del_i\bar A_i^a\del_j\bar A_j^a
-(D_{\tau}^{SU(2)}\bar A_{\tau})^a\del_i\bar A_i^a
\nonumber\\
&+&ig\biggl[\del_i\bar A_i^a
\left(\bar\Phi^{\dagger}{\tau^a\over 2}\vev{\Phi}-
\vev{\Phi}^{\dagger}{\tau^a\over 2}\bar\Phi\right)\biggr]
-{\xi\over 2}(D_{\tau}^{SU(2)}\bar A_{\tau})^a(D_{\tau}^{SU(2)}\bar A_{\tau})^a
\nonumber\\
&+&i\xi g
\biggl[(D_{\tau}^{SU(2)}\bar A_{\tau})^a\left(\bar\Phi^{\dagger}{\tau^a\over 2}\vev{\Phi}-
\vev{\Phi}^{\dagger}{\tau^a\over 2}\bar\Phi\right)\biggr]
+{\xi g^2\over 2}\left(\bar\Phi^{\dagger}{\tau^a\over 2}\vev{\Phi}-
\vev{\Phi}^{\dagger}{\tau^a\over 2}\bar\Phi\right)^2
\nonumber\\
&-&i\bar C^a \del_i^2 C^a
-i\xi \bar C^a (D_{\tau}^{SU(2)}D_{\tau}^{SU(2)}C)^a
\nonumber\\
&+&ig\xi \bar C^a\Bigl({g\over 2}\vev{\Phi}^{\dagger}\vev{\Phi}C^a
+{g_Y\over2 }C\vev{\Phi}^{\dagger}\tau^a\vev{\Phi}\Bigr).
\label{shiki55}
\end{eqnarray}
Likewise we obtain the quadratic terms of the $U(1)_Y$ part,
\begin{eqnarray}
&&{\cal L}_{\rm gf}^{U(1)_Y~(2)}=(-i)\delta_B(CF)\big|_{\rm quadratic }\nonumber\\
&= &-{1\over 2\xi}\del_i\bar B_i\del_j\bar B_j
-\del_i\bar B_i\del_{\tau}\bar B_{\tau}
+{ig_Y\over 2}\biggl[\del_i\bar B_i
\biggl(\bar\Phi^{\dagger}\vev{\Phi}-\vev{\Phi}^{\dagger}\bar\Phi\biggr)\biggr]
-{\xi\over 2}\del_{\tau}\bar B_{\tau}\del_{\tau}\bar B_{\tau}
\nonumber\\
&+&i{\xi g_Y\over 2}
\biggl[\del_{\tau}\bar B_{\tau}\biggl(\bar\Phi^{\dagger}\vev{\Phi}-\vev{\Phi}^{\dagger}\bar\Phi\biggr)
\biggr]
+{\xi g_Y^2\over 8}\biggl(\bar\Phi^{\dagger}\vev{\Phi}-\vev{\Phi}^{\dagger}\bar\Phi\biggr)^2
\nonumber\\
&-&i\bar C \del_i^2 C
-i\xi \bar C \del_{\tau}^2C
+i\xi g_Y C\vev{\Phi}^{\dagger}\biggl(g{\tau^a\over 2}C^a+{g_Y\over 2}C\biggr)\vev{\Phi}.
\label{shiki56}
\end{eqnarray}
The first term in the third line of Eq.(\ref{shiki46}) (the
second term in the third line of Eq. (\ref{shiki46}) ) is canceled by the second 
term in the second line of Eq. (\ref{shiki55}) (the second term in the second line 
of Eq. (\ref{shiki56})) after the partial integration. The third line of 
Eq. (\ref{shiki50}) is canceled by the first term in the 
third line of Eq. (\ref{shiki55}) and the third term in the second line of Eq. (\ref{shiki56})
after the partial integration.

By noting that
\begin{eqnarray}
(\bar D_{\tau}\bar\Phi)^{\dagger}\bar A_{\tau}\vev{\Phi}&=&-\bar\Phi^{\dagger}
(D_{\tau}^{SU(2)}\bar A_{\tau})\vev{\Phi}+ig\bar\Phi^{\dagger}\bar A_{\tau}\vev{A_{\tau}}\vev{\Phi}
+i{g_Y\over 2}\bar\Phi^{\dagger}\vev{B_{\tau}}\bar A_{\tau}\vev{\Phi},
\label{shiki57}\\
(\bar D_{\tau}\bar\Phi)^{\dagger}\bar B_{\tau}\vev{\Phi}&=&
-\bar\Phi^{\dagger}\del_{\tau}\bar B_{\tau}\vev{\Phi}
+ig\bar\Phi^{\dagger}\vev{A_{\tau}}\bar B_{\tau}\vev{\Phi}+i{g_Y\over 2}\bar\Phi^{\dagger}
\vev{B_{\tau}}\bar B_{\tau}\vev{\Phi},
\label{shiki58}
\end{eqnarray}
where the partial integration has been performed, the second line of Eq. (\ref{shiki50}) 
is recast as
\begin{eqnarray}
&&
-(\bar D_{\tau}\bar\Phi)^{\dagger}\left(-ig\bar A_{\tau}\vev{\Phi}-
i{g_Y\over 2}\bar B_{\tau}\vev{\Phi}\right)
-\left(ig\vev{\Phi}^{\dagger}\bar A_{\tau}+
i{g_Y\over 2}\vev{\Phi}^{\dagger}\bar B_{\tau}\right)\bar D_{\tau}\bar \Phi
\nonumber\\
&=&ig\biggl[
-\bar\Phi^{\dagger}
(D_{\tau}^{SU(2)}\bar A_{\tau})\vev{\Phi}+ig\bar\Phi^{\dagger}\bar A_{\tau}\vev{A_{\tau}}\vev{\Phi}
+i{g_Y\over 2}\bar\Phi^{\dagger}\vev{B_{\tau}}\bar A_{\tau}\vev{\Phi}\nonumber\\
&+&\vev{\Phi}^{\dagger}(D_{\tau}^{SU(2)}\bar A_{\tau})\bar\Phi +
ig\vev{\Phi}^{\dagger}\vev{A_{\tau}}\bar A_{\tau}\bar\Phi
+i{g_Y\over 2}\vev{\Phi}^{\dagger}\bar A_{\tau}\vev{B_{\tau}}\bar\Phi
\biggr]\nonumber\\
&+&i{g_Y\over 2}
\biggl[-\bar\Phi^{\dagger}\del_{\tau}\bar B_{\tau}\vev{\Phi}
+ig\bar\Phi^{\dagger}\vev{A_{\tau}}\bar B_{\tau}\vev{\Phi}+i{g_Y\over 2}\bar\Phi^{\dagger}
\vev{B_{\tau}}\bar B_{\tau}\vev{\Phi}\nonumber\\
&+&\vev{\Phi}^{\dagger}\del_{\tau}\bar B_{\tau}\bar\Phi +
ig \vev{\Phi}^{\dagger}\bar B_{\tau}\vev{A_{\tau}}\bar\Phi
+i{g_Y\over 2}\vev{\Phi}^{\dagger}\bar B_{\tau}\vev{B_{\tau}}\bar\Phi
\biggr].
\label{shiki59}
\end{eqnarray}
The first terms in the second and the third line
of Eq. (\ref{shiki59}), together with the 
first term in the fourth line of Eq. (\ref{shiki55}) 
are summarized into the compact form given by the first term in 
Eq. (\ref{shiki60}) below. And likewise the first terms in the fourth and the fifth 
line of Eq. (\ref{shiki59}), together with the first term in the third line of Eq. (\ref{shiki56}) 
are summarized into the compact form given by the second 
term in Eq. (\ref{shiki60}) below. The compact form is given by
\begin{equation}
ig(\xi -1)\Bigl(\bar\Phi^{\dagger}(D_{\tau}^{SU(2)}\bar A_{\tau})\vev{\Phi}
-\vev{\Phi}^{\dagger}(D_{\tau}^{SU(2)}\bar A_{\tau})\bar\Phi
\Bigr)
+{ig_Y\over 2}(\xi - 1)\Bigl(\del_{\tau}\bar B_{\tau}\bar\Phi^{\dagger}\vev{\Phi}
-\vev{\Phi}^{\dagger}\bar\Phi\del_{\tau}\bar B_{\tau}\Bigr),
\label{shiki60}
\end{equation}
which vanishes for the Feynman gauge $\xi =1$. 
%
%
%
%

The quadratic part of the Lagrangian for the gauge fields $\bar A_i^a$ and $\bar B_i$ 
is given by 
\begin{eqnarray}
{\cal L}_{\rm gauge}^{(2)}&=&
{1\over 2}\bar A_i^a\left(\delta_{ij}\del_l^2-\left(1-{1\over \xi}\right)\del_i\del_j\right)\bar A_j^a
-{1\over 2}(D_{\tau}^{SU(2)}\bar A_i)^a(D_{\tau}^{SU(2)}\bar A_i)^a\nonumber\\
&+&{1\over 2}\bar B_i\left(\delta_{ij}\del_l^2-\left(1-{1\over \xi}\right)\del_i\del_j\right)\bar B_j
-{1\over 2}\del_{\tau}\bar B_i\del_{\tau}\bar B_i\nonumber\\
&-&g^2\vev{\Phi}^{\dagger}\bar A_i\bar A_i\vev{\Phi}
-gg_Y\vev{\Phi}^{\dagger}\bar A_i\bar B_i\vev{\Phi}
-\left({g_Y\over 2}\right)^2\vev{\Phi}^{\dagger}\bar B_i\bar B_i\vev{\Phi}.
\label{shiki61}
\end{eqnarray}
The quadratic part for the ghost fields $C^a, \bar C^a, C$ and $\bar C$
is given by
\begin{eqnarray}
{\cal L}_{\rm ghost}^{(2)}
&=&
-i\bar C^a\del_i^2 C^a-i\xi \bar C^a (D_{\tau}^{SU(2)}D_{\tau}^{SU(2)}C)^a
-i\bar C\del_i^2 C-i\xi \bar C\del_{\tau}^2C
\nonumber\\
&+&i\xi g\bar C^a\Bigl({g\over 2}\abs{\vev{\Phi}}^2C^a
+{g_Y\over 2}C\vev{\Phi}^{\dagger}\tau^a\vev{\Phi}\Bigr)
+i\xi g_Y\bar C\vev{\Phi}^{\dagger}\left({g\over 2}\tau^a C^a
+{g_Y\over 2}C\right)\vev{\Phi}.\nonumber\\
\label{shiki62}
\end{eqnarray}
The quadratic part for the scalar fields $\bar A_{\tau}^a, \bar B_{\tau}$ and $\bar\Phi$
is given by
\begin{eqnarray}
{\cal L}_{\rm scalar}^{(2)}
&=&-{1\over 2}\del_i\bar A_{\tau}^a\del_i\bar A_{\tau}^a-
{\xi\over 2}(D_{\tau}^{SU(2)}\bar A_{\tau})^a(D_{\tau}^{SU(2)}\bar A_{\tau})^a
\nonumber\\
&-&{1\over 2}\del_i\bar B_{\tau}\del_i\bar B_{\tau}-{\xi\over 2}
\del_{\tau}\bar B_{\tau}\del_{\tau}\bar B_{\tau}
-\del_i\bar\Phi^{\dagger}\del_i\bar \Phi - 
(\bar D_{\tau}\bar\Phi)^{\dagger}\bar D_{\tau}\bar\Phi
\nonumber\\
&-&g^2\biggl(
2\vev{\Phi}^{\dagger}\vev{A_{\tau}}\bar A_{\tau}\bar\Phi
+2\bar\Phi^{\dagger}\bar A_{\tau}\vev{A_{\tau}}\vev{\Phi}+\vev{\Phi}^{\dagger}
\bar A_{\tau}\bar A_{\tau}\vev{\Phi}\biggr)
\nonumber\\
&-&{gg_Y \over 2}
\biggl(2\bar\Phi^{\dagger}\vev{B_{\tau}}\bar A_{\tau}\vev{\Phi}+
2\vev{\Phi}^{\dagger}\bar A_{\tau}\vev{B_{\tau}}\bar\Phi
+2\bar\Phi^{\dagger}\vev{A_{\tau}}\bar B_{\tau}\vev{\Phi}
\nonumber\\
&&+2\vev{\Phi}^{\dagger}\bar B_{\tau}\vev{A_{\tau}}\bar\Phi 
+2\vev{\Phi}^{\dagger}\bar A_{\tau}\bar B_{\tau}\vev{\Phi}\biggr)
\nonumber\\
&-&\left({g_Y\over 2}\right)^2
\biggl(
2\bar\Phi^{\dagger}\vev{B_{\tau}}\bar B_{\tau}\vev{\Phi}+
2\vev{\Phi}^{\dagger}\bar B_{\tau}\vev{B_{\tau}}\bar\Phi
+\vev{\Phi}^{\dagger}\bar B_{\tau}\bar B_{\tau}\vev{\Phi}
\biggr)\nonumber\\
&+&\mu^2\bar\Phi^{\dagger}\bar\Phi
-{\lambda\over 2}\biggl(2\abs{\vev{\Phi}}^2\abs{\bar\Phi}^2+
2(\vev{\Phi}^{\dagger}\bar\Phi)(\bar\Phi^{\dagger}\vev{\Phi})+
(\vev{\Phi}^{\dagger}\bar\Phi)^2
+(\bar\Phi^{\dagger}\vev{\Phi})^2\biggr)
\nonumber\\
&+&\xi {g^2\over 2}\left({1\over 4}(\bar\Phi^{\dagger}\vev{\Phi})^2
+{1\over 4}(\vev{\Phi}^{\dagger}\bar\Phi)^2+
{1\over 2}(\bar\Phi^{\dagger}\vev{\Phi})(\vev{\Phi}^{\dagger}\bar\Phi)
-\abs{\vev{\Phi}}^2\abs{\bar\Phi}^2\right)\nonumber\\
&+&\xi {g_Y^2\over 8}\biggl((\bar\Phi^{\dagger}\vev{\Phi})^2+(\vev{\Phi}^{\dagger}\bar\Phi)^2
-2(\bar\Phi^{\dagger}\vev{\Phi})(\vev{\Phi}^{\dagger}\bar\Phi)\biggr)\nonumber\\
&+&ig(\xi -1)\biggl(\bar\Phi^{\dagger}(D_{\tau}^{SU(2)}\bar A_{\tau})\vev{\Phi}
-\vev{\Phi}^{\dagger}(D_{\tau}^{SU(2)}\bar A_{\tau})\bar\Phi
\biggr)
\nonumber\\
&+&{ig_Y\over 2}(\xi - 1)\biggl(\del_{\tau}\bar B_{\tau}\bar\Phi^{\dagger}\vev{\Phi}
-\vev{\Phi}^{\dagger}\bar\Phi\del_{\tau}\bar B_{\tau}\biggr).
\label{shiki63}
\end{eqnarray}
The last two lines in Eq. (\ref{shiki63}) vanish when we take the Feynman gauge $\xi=1$.

Let us put the parametrization (\ref{shiki2}) into Eqs. (\ref{shiki61}), (\ref{shiki62}) and (\ref{shiki63}). 
We obtain for the gauge sector ${\cal L}_{\rm gauge}^{(2)}$ that
\begin{eqnarray}
&&{\cal L}_{\rm gauge}^{(2)}\nonumber\\
&=&W_i^-\biggl[\delta_{ij}\del_l^2 -\left(1-{1\over \xi}\right)\del_i\del_j
+\biggl(\Bigl(\del_{\tau}-i(2\pi T)(2\varphi)\Bigr)^2
-{g^2\over 4}(v^2+p^2)\biggr)\delta_{ij}\biggr]W_j^+
\nonumber\\
&+&\half\bar A_i^3\biggl[\delta_{ij}\del_l^2 -\left(1-{1\over \xi}\right)\del_i\del_j
+\biggl(\del_{\tau}^2-{g^2\over 4}(v^2+p^2)\biggr)\delta_{ij}\biggr]\bar A_j^3
\nonumber\\
&+&
\half\bar B_i\biggl[\delta_{ij}\del_l^2 -\left(1-{1\over \xi}\right)\del_i\del_j
+\biggl(\del_{\tau}^2-{g_Y^2\over 4}(v^2+p^2)\biggr)\delta_{ij}\biggr]\bar B_j
\nonumber\\
&-&{gg_Y\over 4}(p^2-v^2)\bar A_i^3\bar B_i
-{\sqrt{2}\over 4}gg_Ypv(W_i^+ + W_i^-)\bar B_i 
\label{shiki64}\\
&=&
(W_i^-, ~\bar A_i^3,~ \bar B_i)
M_{\rm gauge}^2
\begin{pmatrix}
W_j^+\\[0.2cm]
\bar A_j^3\\[0.2cm]
\bar B_j
\end{pmatrix},
\label{shiki65}
\end{eqnarray}
where
\begin{equation}
M^2_{\rm gauge}\equiv 
\begin{pmatrix}
D^{W^{\pm}}_{ij}& 0& -{\sqrt{2}\over 4}gg_Ypv\delta_{ij}\\
0&\half D^{A^3}_{ij}& -\half {1\over 4}gg_Y(p^2-v^2)\delta_{ij} \\
 -{\sqrt{2}\over 4}gg_Ypv\delta_{ij} & -\half {1\over 4}gg_Y(p^2-v^2)\delta_{ij} 
 &\half D^B_{ij}
\end{pmatrix}.
\label{shiki66}
\end{equation}
Here we have defined 
\begin{equation}
W_{\mu}^{\pm}\equiv {1\over\sqrt{2}}\left(\bar A_{\mu}^1 \mp i\bar A_{\mu}^2\right)\quad (\mu=\tau, 1, 2, 3)
\label{shiki67}
\end{equation}
and
\begin{eqnarray}
D^{W^{\pm}}_{ij}&=&\delta_{ij}\del_l^2 -\left(1-{1\over \xi}\right)\del_i\del_j
+\biggl(\Bigl(\del_{\tau}-i(2\pi T)(2\varphi)\Bigr)^2-{g^2\over 4}(v^2+p^2)\biggr)\delta_{ij},
\nonumber\\
D^{A^3}_{ij}&=&\delta_{ij}\del_l^2 -\left(1-{1\over \xi}\right)\del_i\del_j
+\biggl(\del_{\tau}^2-{g^2\over 4}(v^2+p^2)\biggr)\delta_{ij},
\nonumber\\
D^{B}_{ij}&=&\delta_{ij}\del_l^2 -\left(1-{1\over \xi}\right)\del_i\del_j
+\biggl(\del_{\tau}^2-{g_Y^2\over 4}(v^2+p^2)\biggr)\delta_{ij} .
\label{shiki68}
\end{eqnarray}
One observes that there is a coupling between $\bar W_i$ and $\bar B_j$ that
breaks the $U(1)_{\rm em}$ invariance due to the vacuum 
expectation value $p$. This also happens in the scalar, ghost and fermion sectors 
discussed below.
The scalar sector is given by
\begin{eqnarray}
{\cal L}_{\rm scalar}^{(2)}
&=&
W_{\tau}^- \biggl[\del_i^2+\xi \Bigl(\del_{\tau}-i(2\pi T)(2\varphi)\Bigr)^2-{g^2\over 4}(p^2+v^2)\biggr]W_{\tau}^+
\nonumber\\
&+&\half\bar A_{\tau}^3\biggl[\del_i^2+\xi \del_{\tau}^2-{g^2\over 4}
(v^2+p^2)\biggr]\bar A_{\tau}^3+
\half\bar B_{\tau}\biggl[\del_i^2+\xi \del_{\tau}^2-{g_Y^2\over 4}
(v^2+p^2)\biggr]\bar B_{\tau}
\nonumber \\
&+&
\half h \biggl[\del_i^2+\Bigl(\del_{\tau}-i(2\pi T)\bigl(-\varphi +{\theta\over 2}\bigr)\Bigr)^2
+\mu^2-\lambda \left({3\over 2}v^2 
+{p^2\over 2}\right)-{\xi\over 4}g^2p^2\biggr]h
\nonumber\\
&+&
\half G^0 \biggl[\del_i^2+\Bigl(\del_{\tau}-i(2\pi T)\bigl(-\varphi +{\theta\over 2}\bigr)\Bigr)^2
+\mu^2-\lambda \left({v^2\over 2} 
+{p^2\over 2}\right)-\xi {g^2+g_Y^2\over 4}v^2-\xi {g^2\over 4}p^2\biggr]G^0
\nonumber\\
&+&G^- \biggl[\del_i^2+\Bigl(\del_{\tau}-i(2\pi T)\bigl(\varphi +{\theta\over 2}\bigr)\Bigr)^2
+\mu^2-\lambda \left({v^2\over 2} 
+p^2\right)-\xi {g^2\over 4}v^2-\xi {g^2+g_Y^2\over 8}p^2\biggr]G^+
\nonumber\\
&-&{\lambda\over 2}\biggl[{p^2\over 2}\left({G^+}^2+{G^-}^2\right)+\sqrt{2}pv (G^+ + G^-)h\biggr]
+\xi{g^2+g_Y^2\over 16}p^2({G^+}^2 + {G^-}^2)
\nonumber\\
&-&g(2\pi T)
\biggl[
{p\over\sqrt{2}}\left(\varphi +{\theta\over 2}\right)\Bigl\{(G^+ + G^-)\bar A_{\tau}^3
+(W_{\tau}^+ + W_{\tau}^-)h
\nonumber\\
&+&i(W_{\tau}^+- W_{\tau}^-)G^0\Bigr\}
-v\left(\varphi -{\theta\over 2}\right)\Bigl(W_{\tau}^-G^+ + W_{\tau}^+ G^- -\bar A_{\tau}^3 h\Big)
\biggr]
\nonumber\\
&-&g_Y(2\pi T)
\biggl[
{p\over\sqrt{2}}\left(\varphi +{\theta\over 2}\right)(G^+ + G^-)\bar B_{\tau}
-v\left(\varphi -{\theta\over 2}\right)\bar B_{\tau} h
\biggr]
\nonumber\\
&-&gg_Y\biggl[
{1\over 4}(p^2 - v^2)\bar A_{\tau}^3 \bar B_{\tau}+{\sqrt{2}\over 4}pv (W_{\tau}^+ + W_{\tau}^- )\bar B_{\tau}
\biggr]
\nonumber\\
&+&\xi{\sqrt{2}\over 8}g^2pv (G^+ + G^-)h +\xi {\sqrt{2}\over 8}g_Y^2 pv (G^+ - G^-)iG^0
\nonumber\\
&+&ig(\xi -1){1\over2\sqrt{2}}\biggl[
p(G^- - G^+)\del_{\tau}\bar A_{\tau}^3\nonumber\\
&+&\sqrt{2} v\Bigl(G^- (\del_{\tau}-i(2\pi T)(2\varphi))W_{\tau}^+ 
- G^+ (\del_{\tau}-i(2\pi T)(-2\varphi))W_{\tau}^-\Bigr)
\nonumber\\
&+&p\Bigl(
(\del_{\tau}-i(2\pi T)(-2\varphi))W_{\tau}^- - 
(\del_{\tau}-i(2\pi T)(2\varphi))W_{\tau}^+
\Bigr)h \nonumber\\
&-& ip\Bigl(
(\del_{\tau}-i(2\pi T)(-2\varphi))
W_{\tau}^- 
+ 
(\del_{\tau}-i(2\pi T)(2\varphi))
W_{\tau}^+ \Bigr)G^0 
+i\sqrt{2} v G^0\del_{\tau}\bar A_{\tau}^3
\biggr]\nonumber\\
&+&ig_Y(\xi -1) {1\over2\sqrt{2}}\biggl[
p(G^- - G^+)\del_{\tau}\bar B_{\tau} -\sqrt{2} i v G^0 \del_{\tau}\bar B_{\tau}
\biggr].
\label{shiki69}
\end{eqnarray}
Here we have defined 
$$
\bar \Phi=\begin{pmatrix}
G^+\\
{1\over\sqrt{2}}(h+iG^0)
\end{pmatrix}.
$$
In terms of the real fields defined by Eq. (\ref{shiki67}) and
\begin{equation}
%
%
G^{\pm}\equiv {1\over\sqrt{2}}(g^1 \mp ig^2),
\label{shiki70}
\end{equation} 
Eq.  (\ref{shiki69}) becomes
\begin{eqnarray}
&&{\cal L}_{\rm scalar}^{(2)}\nonumber\\
&=&\half \bar A_{\tau}^1 \biggl[\del_i^2+\xi (\del_{\tau}^2 -(2\pi T)^2(2\varphi)^2)-
{g^2\over 4}(p^2+v^2)\biggr]\bar A_{\tau}^1
\nonumber\\
&+&\half \bar A_{\tau}^2\biggl[\del_i^2+\xi (\del_{\tau}^2 -(2\pi T)^2(2\varphi)^2)-
{g^2\over 4}(p^2+v^2)\biggr]\bar A_{\tau}^2
-2\pi T(2\varphi)(\bar A_{\tau}^1\del_{\tau}\bar A_{\tau}^2 - \bar A_{\tau}^2\del_{\tau}\bar A_{\tau}^1)
\nonumber\\
&+&\half\bar A_{\tau}^3\biggl[\del_i^2+\xi \del_{\tau}^2-{g^2\over 4}
(v^2+p^2)\biggr]\bar A_{\tau}^3
+
\half\bar B_{\tau}\biggl[\del_i^2+\xi \del_{\tau}^2-{g_Y^2\over 4}
(v^2+p^2)\biggr]\bar B_{\tau}
\nonumber\\
&+&
\half h \biggl[\del_i^2+\del_{\tau}^2 - (2\pi T)^2\Bigl(-\varphi+{\theta\over 2}\Bigr)^2
+\mu^2-\lambda \left({3\over 2}v^2 
+{p^2\over 2}\right)-{\xi\over 4}g^2p^2\biggr]h
\nonumber\\
&+&
\half G^0 \biggl[\del_i^2+\del_{\tau}^2 -(2\pi T)^2\Bigl(-\varphi+{\theta\over 2}\Bigr)^2
+\mu^2-\lambda \left({v^2\over 2} 
+{p^2\over 2}\right)-\xi {g^2+g_Y^2\over 4}v^2-\xi {g^2\over 4}p^2\biggr]G^0
\nonumber\\
&+&\half g^1 \biggl[\del_i^2+\del_{\tau}^2 -(2\pi T)^2\Bigl(\varphi+{\theta\over 2}\Bigr)^2+\mu^2
-{\lambda\over 2}v^2-{3 \over 2}\lambda p^2-\xi {g^2\over 4}v^2\biggr]g^1
\nonumber\\
&+&\half g^2 \biggl[\del_i^2+\del_{\tau}^2 -(2\pi T)^2\Bigl(\varphi+{\theta\over 2}\Bigr)^2+\mu^2
-{\lambda\over 2}v^2-{\lambda\over 2} p^2-\xi {g^2\over 4}v^2-\xi {g^2+g_Y^2\over 4}p^2\biggr]g^2
\nonumber\\
&-&2\pi T\Bigl(\varphi+ {\theta\over 2}\Bigr)(g^1\del_{\tau}g^2 -g^2\del_{\tau}g^1)
\nonumber\\
&-&g(2\pi T)
\biggl[p\Bigl(\varphi+{\theta\over 2}\Bigr)(g^1\bar A_{\tau}^3 +\bar A_{\tau}^1 h +\bar A_{\tau}^2 G^0)
-v\Bigl(\varphi-{\theta\over 2}\Bigr)(\bar A_{\tau}^1g^1 +\bar A_{\tau}^2 g^2 -\bar A_{\tau}^3 h)
\biggr]
\nonumber\\
&-&g_Y(2\pi T)
\biggl[p\Bigl(\varphi+{\theta\over 2}\Bigr)g^1 \bar B_{\tau} -
v\Bigl(\varphi -{\theta\over 2}\Bigr)\bar B_{\tau} h\biggr]
\nonumber\\
&-&gg_Y\biggl[
{1\over 4}(p^2 - v^2)\bar A_{\tau}^3 \bar B_{\tau}+{1\over 2}pv\bar A_{\tau}^1\bar B_{\tau}
\biggr]
+\xi{g^2\over 4}pv g^1h 
+\xi {g_Y^2\over 4} pv g^2G^0-\lambda pv hg^1\nonumber\\
&-&{g\over 2}(\xi -1) 
\biggl[
(pg^2+vG^0)\del_{\tau}\bar A_{\tau}^3 -v (g^1\del_{\tau}\bar A_{\tau}^2 - g^2 \del_{\tau}\bar A_{\tau}^1)
+p (h\del_{\tau}\bar A_{\tau}^2 -G^0\del_{\tau}\bar A_{\tau}^1)\nonumber\\
&-&(2\pi T)(2\varphi)v (g^1\bar A_{\tau}^1 + g^2 \bar A_{\tau}^2) 
+(2\pi T)(2\varphi) p (h\bar A_{\tau}^1+ G^0 \bar A_{\tau}^2) 
\biggr]\nonumber\\
&-&{g_Y\over 2}(\xi -1) (pg^2 -v G^0)\del_{\tau}\bar B_{\tau}
\label{shiki71}\\
&=&\half
(\bar A_{\tau}^1,~\bar A_{\tau}^2,~g^1,~ g^2,~\bar A_{\tau}^3,~ \bar B_{\tau},~ h,~G^0)
M_{\rm scalar}^2
\begin{pmatrix}
\bar A_{\tau}^1\\
\bar A_{\tau}^2\\
g^1\\
g^2\\
\bar A_{\tau}^3\\
\bar B_{\tau}\\
h\\
G^0
\end{pmatrix},
\label{shiki72}
\end{eqnarray}
where $M_{\rm scalar}^2$ is defined by 
\begin{equation}
M_{\rm scalar}^2=\begin{pmatrix}
A & a & b &-\bar c & 0 & c & d & \bar d \\
-a &A & \bar c& b &0  &0  &-\bar d & d \\
b    &  -\bar c & B &g & h & i & j & 0 \\
 \bar c  &b   & -g & C & \bar d & \bar b & 0 & k \\
0   &  0 & h &  -\bar d  & D & l & m & -\bar c\\
 c  & 0  & i & -\bar b   & l  & E & n & \bar e \\
 d   & \bar d  &j  & 0   &  m &n  & F & 0 \\
 -\bar d    & d & 0 & k   &  \bar c &  -\bar e&0  & G \\
\end{pmatrix}.
\label{shiki73}
\end{equation}
The components in $M_{\rm scalar}^2$ are given as
\begin{eqnarray}
A&=&\del_i^2 +\xi \left(\del_{\tau}^2 -(2\pi T)^2 (2\varphi)^2\right)-{g^2\over 4}(p^2+v^2),
\nonumber\\
B&=&\del_i^2 +\del_{\tau}^2 -(2\pi T)^2 \left(\varphi+{\theta\over 2}\right)^2+\mu^2
-{\lambda\over 2}v^2 -{3\over 2}\lambda p^2 -\xi {g^2\over 4}v^2, 
\nonumber\\
C&=&\del_i^2 +\del_{\tau}^2 -(2\pi T)^2\left(\varphi+{\theta\over 2}\right)^2+\mu^2
-{\lambda\over 2}v^2-{\lambda\over 2}p^2  -\xi {g^2\over 4}v^2-{\xi\over 4}(g^2+g_Y^2)p^2, 
\nonumber\\
D&=&\del_i^2 +\xi \del_{\tau}^2-{g^2\over 4}(p^2+v^2), 
\nonumber\\
E&=&\del_i^2 +\xi \del_{\tau}^2-{g_Y^2\over 4}(p^2+v^2), 
\nonumber\\
F&=&\del_i^2 +\del_{\tau}^2-(2\pi T)^2\left(-\varphi +{\theta\over 2}\right)^2+\mu^2 -
\lambda \left({3\over 2}v^2+{p^2\over 2}\right)-{\xi\over 4}g^2p^2, 
\nonumber\\
G&=&\del_i^2 +\del_{\tau}^2 -(2\pi T)^2\left(-\varphi+{\theta\over 2}\right)^2
+\mu^2-\lambda \left({v^2\over 2}+{p^2\over 2}\right)-\xi{g^2+g_Y^2\over 4}v^2-\xi{g^2\over 4}p^2,
\nonumber\\
a&=&-2(2\pi T)(2\varphi)\del_{\tau},\quad
b=gv(2\pi T)\Bigl(\varphi-{\theta\over 2}\Bigr)- \eta_2 (2\pi T)(2\varphi)v,
\nonumber\\
c&=&-{gg_Y\over 2}pv,\quad
d=-gp(2\pi T)\Bigl(\varphi+{\theta\over 2}\Bigr)+\eta_2 (2\pi T)(2\varphi) p,
\nonumber\\
g&=&-2(2\pi T)\left(\varphi+{\theta\over 2}\right)\del_{\tau}, ~
h=-gp(2\pi T)\Bigl(\varphi +{\theta\over 2}\Bigr),~
i=-g_Yp(2\pi T)\Bigl(\varphi+{\theta\over 2}\Bigr),
\nonumber\\
j&=&\xi {g^2\over 4}pv-\lambda pv, ~~
k=\xi {g_Y^2\over 4}pv,~ l=-{gg_Y\over 4}(p^2 - v^2),~
m=-gv (2\pi T)\Bigl(\varphi -{\theta\over 2}\Bigr),
\nonumber\\
n&=&g_Yv (2\pi T)\Bigl(\varphi -{\theta\over 2}\Bigr),\quad 
\bar c=\eta_2 v\del_{\tau},\quad \bar d=\eta_2 p\del_{\tau},\quad
\bar b=\eta_1 p\del_{\tau},\quad
\bar e=\eta_1 v\del_{\tau},
\label{shiki74}
\end{eqnarray}
where we have defined $\eta_2\equiv -g(\xi -1)/2, \eta_1\equiv -g_Y(\xi -1)/2$, which 
vanishes if we take the Feynman gauge $\xi=1$.

Let us proceed to the ghost sector. As usual, it is convenient to introduce
\begin{equation}
C^{\pm}\equiv {1\over\sqrt{2}}(C^1 \mp iC^2),~~
{\bar C}^{\pm}\equiv {1\over\sqrt{2}}(\bar C^1 \mp iC^2).
\label{shiki75}
\end{equation}
Then the ghost sector is given by
\begin{eqnarray}
{\cal L}_{\rm ghost}^{(2)}
&=&i\bar C^+\biggl(-\del_i^2 -\xi \Bigl(\del_{\tau}-i(2\pi T)(2\varphi)\Bigr)^2+\xi{g^2\over 4}(p^2+v^2)\biggr)C^- 
\nonumber\\
&+&i\bar C^-\left(-\del_i^2 -\xi \Bigl(\del_{\tau}-i(2\pi T)(2\varphi)\Bigr)^2+\xi{g^2\over 4}(p^2+v^2)\right)C^+ 
\nonumber\\
&+&i\bar C^3\left(-\del_i^2 - \xi\del_{\tau}^2+\xi {g^2\over 4}(p^2+v^2)\right)C^3
\nonumber\\
&+&i\bar C\left(-\del_i^2 -\xi \del_{\tau}^2+\xi {g_Y^2\over 4}(p^2+v^2)\right)C
\nonumber\\
&-&i\xi {gg_Y\over 4}(p^2 - v^2)C \bar C^3+i\xi {gg_Y\over 4}(p^2 - v^2)\bar C C^3
\nonumber\\
&-&i\xi {\sqrt{2}\over 4}gg_Y pv 
\biggl[C\left (\bar C^+ + \bar C^-\right )- \bar C\left(C^+ + C^-\right)\biggr]
\label{shiki76}\\
&=&
i(\bar C^+,~ \bar C^-,~\bar C^3,~\bar C)
M_{\rm ghost}^2
\begin{pmatrix}
C^-\\
C^+\\
C^3\\
C 
\end{pmatrix},
\label{shiki77}
\end{eqnarray}
where $M_{\rm ghost}^2$ is given by
\begin{eqnarray}
M_{\rm ghost}^2=\begin{pmatrix}
-D^{W^{\pm}} & 0 & 0 & \xi {\sqrt{2}\over 4}gg_Ypv \\
0 & -D^{W^{\pm}} & 0 &\xi {\sqrt{2}\over 4}gg_Ypv \\
0&0 & -D^{A^3}& \xi {gg_Y\over 4}(p^2-v^2) \\
 \xi {\sqrt{2}\over 4}gg_Ypv   
 &\xi {\sqrt{2}\over 4}gg_Ypv  & \xi {gg_Y\over 4}(p^2-v^2)& -D^B\\
\end{pmatrix}.
\label{shiki78}
\end{eqnarray}
Here we have defined 
\begin{eqnarray}
D^{W^{\pm}}&=&\del_i^2+\xi\Bigl[\Bigl(\del_{\tau}-i(2\pi T)(2\varphi)\Bigr)^2-{g^2\over 4}(p^2+v^2)\Bigr],
\nonumber\\
D^{A^3}&=&\del_i^2+\xi \Bigl(\del_{\tau}^2- {g^2\over 4}(p^2+v^2)\Bigr),
\nonumber\\
D^B&=&\del_i^2+\xi\Bigl(\del_{\tau}^2-{g_Y^2\over 4}(p^2+v^2)\Bigr).
\label{shiki79}
\end{eqnarray}
These are the same as (\ref{shiki68}) with $\xi=1$ aside form the factor $\delta_{ij}$.

Let us introduce the fermions of the third generation,
\begin{equation}
Q_L=\begin{pmatrix} t_L \\ b_L\end{pmatrix},~t_R, ~b_R,~l
_L=\begin{pmatrix} \nu^{\tau}_ L \\ \tau_L\end{pmatrix},~\tau_R .
\label{shiki80}
\end{equation}
Since our results do not change even if we introduce the first and second
generations with the mixings among the generations, we consider only the
third generation. We simply assume that the neutrino is massless.
%
%
The Lagrangian for the fermions is
\begin{eqnarray}
{\cal L_{\rm fermion}}&=&
{\bar Q}_L i \gamma^{\mu}D_{\mu}^Q Q_L +{\bar t}_R i \gamma^{\mu}D_{\mu}^{t_R} t_R
+{\bar b}_R i \gamma^{\mu}D_{\mu}^{b_R} b_R \nonumber\\
&+&f_t\left({\bar t}_R {\tilde \Phi}^{\dagger}Q_L + {\bar Q}_L \tilde\Phi t_R\right)
+f_b\left({\bar b}_R \Phi^{\dagger}Q_L + {\bar Q}_L \Phi b_R\right)
\nonumber\\
&+&{\bar l}_L i \gamma^{\mu}D_{\mu}^l l_L + {\bar \tau}_R i \gamma^{\mu}D_{\mu}^{\tau_R} {\tau}_R
+f_{\tau}\left({\bar \tau}_R \Phi^{\dagger}l_L + {\bar l}_L \Phi {\tau}_R\right),
\label{shiki81}
\end{eqnarray}
where the covariant derivatives are given by
\begin{eqnarray}
D_{\mu}^Q &=&\nabla_{\mu}- ig A_{\mu}-i{g_Y\over 2}{1\over 3}B_{\mu}, ~
D_{\mu}^{t_R} =\nabla_{\mu}-i{g_Y\over 2}{4\over 3}B_{\mu},~
D_{\mu}^{b_R} =\nabla_{\mu}-i{g_Y\over 2}\bigl(-{2\over 3}\bigr)B_{\mu},\nonumber\\
D_{\mu}^l &=&\del_{\mu}- ig A_{\mu}-i{g_Y\over 2}(-1)B_{\mu},~
D_{\mu}^{\tau_R} =\del_{\mu}-i{g_Y\over 2}(-2)B_{\mu}
\label{shiki82}
\end{eqnarray}
and $f_{t, b, \tau}$ stands for the Yukawa couplings for the top, bottom and tau. 
Here $\nabla_{\mu}$ stands for the covariant derivative for the $SU(3)_c$, which is defined by
\begin{equation}
\nabla_{\mu}\equiv \del_{\mu} - i g_s G_{\mu}.
\label{shiki83}
\end{equation}
The Euclidean component of the gamma matrices is defined by 
$\gamma_{\tau}\equiv i\gamma_0$ with
\begin{equation}
%
%
\{\gamma_{\mu},~\gamma_{\nu}\}=-2\delta_{\mu\nu}~~(\mu, \nu=\tau, 1, 2, 3).
\label{shiki84}
\end{equation}
Then the quadratic terms from the fermions are
\begin{eqnarray}
{\cal L}_{\rm fermion}^{(2)}
&=&
-i \bar Q_L\biggl(\gamma_{\tau}
\Bigl(\del_{\tau} - ig_s\vev{G_{\tau}} - ig\vev{A_{\tau}}-i{g_Y\over 6}\vev{B_{\tau}}\Bigr)
+\gamma_i \del_i\biggr)Q_L\nonumber\\
&-&i\bar t_R\biggl(\gamma_{\tau}\Bigl(\del_{\tau}- 
ig_s\vev{G_{\tau}}-i{2g_Y\over 3}\vev{B_{\tau}}\Bigr)+\gamma_i \del_i\biggr)t_R
\nonumber\\
&+&f_t\left(\bar t_R\vev{\tilde\Phi}^{\dagger}Q_L + \bar Q_L\vev{\tilde \Phi}t_R\right) \nonumber\\
&-&i\bar b_R\biggl(\gamma_{\tau}\Bigl(\del_{\tau}- 
ig_s\vev{G_{\tau}}+i{g_Y\over 3}\vev{B_{\tau}}\Bigr)+\gamma_i \del_i\biggr)b_R
\nonumber\\
&+&f_b\left(\bar b_R\vev{\Phi}^{\dagger}Q_L + \bar Q_L\vev{\Phi}b_R\right) \nonumber\\
&-&i \bar l_L\biggl(\gamma_{\tau}\Bigl(\del_{\tau} -ig\vev{A_{\tau}}+i{g_Y\over 2}\vev{B_{\tau}}\Bigr)+
\gamma_i \del_i\biggr)l_L
\nonumber\\
&-&i\bar \tau_R\biggl(\gamma_{\tau}\Bigl(\del_{\tau}+ig_Y\vev{B_{\tau}}\Bigr)+\gamma_i \del_i\biggr)\tau_R
+f_{\tau}\left(\bar \tau_R\vev{\Phi}^{\dagger}l_L + \bar l_L\vev{\Phi}\tau_R\right).
\label{shiki85}
\end{eqnarray}
By substituting the parametrizations for the vacuum expectation values (\ref{shiki2}) and 
(\ref{shiki6}) in Eq. (\ref{shiki85}), we obtain that
\begin{eqnarray}
{\cal L}_{\rm fermion}=
(\bar t_L, \bar t_R, \bar b_L, \bar b_R)M_{\rm quark}
\begin{pmatrix}
t_L\\
t_R\\
b_L\\
b_R
\end{pmatrix}
+({\bar\nu}_L, \bar \tau_L, \bar \tau_R,)M_{\rm lepton}
\begin{pmatrix}
\nu_L\\
\tau_L\\
\tau_R
\end{pmatrix},
\label{shiki86}
\end{eqnarray}
where we have defined 
\begin{eqnarray}
M_{\rm quark}&=&
\begin{pmatrix}
-iD_{t_L}& {f_t\over\sqrt{2}}v & 0 & {f_b\over\sqrt{2}}p\\
{f_t\over\sqrt{2}}v  & -iD_{t_R} & {-f_t\over\sqrt{2}}p & 0 \\
0&{-f_t\over\sqrt{2}}p & -iD_{b_L} & {f_b\over\sqrt{2}}v \\
{f_b\over\sqrt{2}}p & 0 & {f_b\over\sqrt{2}}v &-iD_{b_R}  \\
\end{pmatrix},
\label{shiki87}\\
M_{\rm lepton}&=&
\begin{pmatrix}
-iD_{\nu_L}& 0 & {f_{\tau}\over\sqrt{2}}p \\
0  &-i D_{\tau_L} & {f_{\tau}\over\sqrt{2}}v  \\
{f_{\tau}\over \sqrt{2}}p&{f_{\tau}\over\sqrt{2}}v & -iD_{{\tau}_R}  
\label{shiki88}\\
\end{pmatrix}.
\end{eqnarray}
The diagonal elements are given by
\begin{eqnarray}
D_{t_L}&\equiv &\gamma_{\tau}\Bigl(\del_{\tau}-i(2\pi T)\bigl(\omega_r +\varphi+{\theta\over 6}\bigl)\Bigr)+\gamma_i\del_i,
D_{t_R}\equiv \gamma_{\tau}\Bigl(\del_{\tau}-i(2\pi T)\bigl(\omega_r +{2\over 3}\theta\bigr)\Bigr)+\gamma_i\del_i,\nonumber\\
D_{b_L}&\equiv &\gamma_{\tau}\Bigl(\del_{\tau}-i(2\pi T)\bigl(\omega_r -\varphi +{\theta\over 6}\bigr)\Bigr)+\gamma_i\del_i,
D_{b_R}\equiv \gamma_{\tau}\Bigl(\del_{\tau}-i(2\pi T)\bigl(\omega_r-{\theta\over 3}\bigr)\Bigr)+\gamma_i\del_i,\nonumber\\
D_{\tau_L}&\equiv &\gamma_{\tau}\Bigl(\del_{\tau}+i(2\pi T)\bigl(\varphi+{\theta\over 2}\bigr)\Bigr)+\gamma_i\del_i,
D_{\tau_R}\equiv \gamma_{\tau}\Bigl(\del_{\tau}+i(2\pi T)\theta\Bigr)+\gamma_i\del_i,\nonumber\\
D_{\nu_L}&\equiv &\gamma_{\tau}\Bigl(\del_{\tau}-i(2\pi T)\bigl(\varphi-{\theta\over 2}\bigr)\Bigr)+\gamma_i\del_i.
\label{shiki89}
\end{eqnarray}
%
%
Let us finally consider the $SU(3)_c$ gauge sector whose 
Lagrangian, including the gauge fixing and the ghost, is given by
\begin{eqnarray}
{\cal L}_{SU(3)_c}=-\half {\rm tr}\left(G_{\mu\nu}G^{\mu\nu}\right)
-i\delta_B\left(\bar C_s^{\alpha} F_s^{\alpha}\right),
\label{shiki90}
\end{eqnarray}
where $\alpha (=1\sim 8)$ is the $SU(3)_c$ color index and the gauge fixing 
function $F_s^{\alpha}$ is chosen to be
\begin{equation}
F_s^{\alpha}\equiv -\del_i\bar G_i^{\alpha}-\xi_s(D_{\tau}^{SU(3)_c}\bar G_{\tau})^{\alpha}
+{\xi_s\over 2}b_s^{\alpha},
\label{shiki91}
\end{equation}
as usual. The calculations are straightforward and go in 
parallel with the $SU(2)_L$ case except that 
there is no scalar field like the Higgs field in this sector.  Expanding the gauge field $G_{\mu}$ 
around the background $\vev{G_{\tau}}$ and taking the quadratic terms with respect to the 
fluctuations, one obtains that
\begin{eqnarray}
{\cal L}_{SU(3)_c}^{(2)}&=&
{1\over 2}\bar G_i^{\alpha}\left(\delta_{ij}\del_l^2-\left(1-{1\over \xi}\right)\del_i\del_j\right)\bar G_j^{\alpha}
-{1\over 2}(D_{\tau}^{SU(3)_c}\bar G_i)^{\alpha}(D_{\tau}^{SU(3)_c}\bar G_i)^{\alpha}\nonumber\\
&-&{1\over 2}\del_i\bar G_{\tau}^{\alpha}\del_i\bar G_{\tau}^{\alpha}
-{\xi_s\over 2}(D_{\tau}^{SU(3)_c}\bar G_{\tau})^{\alpha}(D_{\tau}^{SU(3)_c}\bar G_{\tau})^{\alpha}\nonumber\\
&-&i\bar C_s^{\alpha}\del_i^2C_s^{\alpha} -i\xi_s\bar C_s^{\alpha}(D_{\tau}^{SU(3)_c}D_{\tau}^{SU(3)_c}C_s)^{\alpha},
\label{shiki92}
\end{eqnarray}
where the covariant derivative in Eqs. (\ref{shiki91}) and (\ref{shiki92}) is defined by
\begin{equation}
D_{\tau}^{SU(3)_c}\bar G_{\mu}\equiv \del_{\tau}\bar G_{\mu} -ig_s[\vev{G_{\tau}},~\bar G_{\mu}]
\quad (\mu=\tau, 1, 2, 3) .
\label{shiki93}
\end{equation}
\begin{flushleft}
{\bf Quadratic terms under the ansatz}
\end{flushleft}
We have obtained the quadratic terms. It is, however, difficult to sum up all the Matsubara
modes $n$ because of the complex dependence on $n$ in the matrices 
(\ref{shiki66}), (\ref{shiki73}), (\ref{shiki78}), (\ref{shiki87}) and (\ref{shiki88}) 
\footnote{The derivative $\del_{\tau}$ is replaced by $i(2\pi T)n~(i(2\pi T)(n+\half))$ in
the momentum space for bosons (fermions).}.
As explained in the text, it may 
be natural to impose the ansatz (\ref{shiki15}) in order to study the effective potential at the one-loop 
level as analytically as possible. Under the ansatz with the Feynman gauge 
$\xi =1$ and $\xi_s=1$, the matrices become so simple that we can diagonalize 
them and sum up all the Matsubara modes.

The quadratic terms under the ansatz for the gauge sector is simplified as
\begin{eqnarray}
&&(W_i^-, \bar A_i^3, \bar B_i)
M^2_{\rm gauge}\big|_{\rm ansatz}
\begin{pmatrix}
W_j^+\\[0.2cm]
\bar A_j^3\\[0.2cm]
\bar B_i
\end{pmatrix}
\nonumber\\
&=&
W_i^- \bar D^{W^{\pm}}\delta_{ij} W_j^+
+\half (\bar A_i^3, \bar B_i)
\begin{pmatrix}
\bar D^{A^3}&  {1\over 4}gg_Yv^2 \\
{1\over 4}gg_Yv^2 &\bar D^B
\end{pmatrix}
\delta_{ij}
\begin{pmatrix}
\bar A_j^3\\
\bar B_j
\end{pmatrix}
\label{shiki94}
\end{eqnarray}
where
\begin{eqnarray}
\bar D^{W^{\pm}}&=&\del_i^2
+\Bigl(\del_{\tau}-i(2\pi T)(2\varphi)\Bigr)^2-{g^2\over 4}v^2,
\nonumber\\
\bar D^{A^3}&=&\del_i^2
+\del_{\tau}^2-{g^2\over 4}v^2,\quad
\bar D^{B}=\del_i^2 +\del_{\tau}^2-{g_Y^2\over 4}v^2.
\label{shiki95}
\end{eqnarray}
Diagonalization of the $\bar A^3_i$ and $\bar B_i$ sector can be done by 
the usual rotation,
\begin{equation}
\begin{pmatrix}
Z_i\\
A_i^{\gamma}
\end{pmatrix}
=
\begin{pmatrix}
c_w& -s_w\\
s_w &c_w
\end{pmatrix}
\begin{pmatrix}
\bar A_i^3\\
\bar B_i
\end{pmatrix}
~\mbox{with}~
\left\{\begin{array}{l}
c_w\equiv \cos\theta_w\equiv {g\over \sqrt{g^2+g_Y^2}},\\
s_w\equiv \sin\theta_w\equiv {g_Y\over\sqrt{g^2+g_Y^2}}.
\end{array}\right.
\label{shiki96}
\end{equation}
Then the eigenvalues in the momentum space are given by
\begin{eqnarray}
-k_i^2-(2\pi T)^2(n-2\varphi)^2-{g^2\over 4}v^2&\cdots &W_i^{\pm},\nonumber\\
-k_i^2-(2\pi T)^2n^2-{g^2+g_Y^2\over 4}v^2&\cdots &Z_i,\nonumber\\
-k_i^2-(2\pi T)^2n^2 &\cdots& A_i^{\gamma}.
\label{shiki97}
\end{eqnarray}
%
%
%
The quadratic terms for the scalar sector are
\begin{eqnarray}
&&\half 
(\bar A_{\tau}^1,~\bar A_{\tau}^2,~g^1,~ g^2,~\bar A_{\tau}^3,~ \bar B_{\tau},~ h,~G^0)
M_{\rm scalar}^2\big|_{\rm ansatz}
\begin{pmatrix}
\bar A_{\tau}^1\\[0.1cm]
\bar A_{\tau}^2\\[0.1cm]
g^1\\[0.1cm]
g^2\\[0.1cm]
\bar A_{\tau}^3\\[0.1cm]
\bar B_{\tau}\\
h\\
G^0
\end{pmatrix}
\label{shiki98}\\
&=&
\half (\bar A_{\tau}^1, \bar A_{\tau}^2)
\begin{pmatrix}
\bar A& a  \\
-a &\bar A\\
\end{pmatrix}
\begin{pmatrix}
\bar A_{\tau}^1\\
\bar A_{\tau}^2
\end{pmatrix}
+
\half  (g^1, g^2)
\begin{pmatrix}
\bar B& \bar g  \\
-\bar g &\bar C\\
\end{pmatrix}
\begin{pmatrix}
g^1\\
g^2
\end{pmatrix}\nonumber\\
&+&
\half  (\bar A_{\tau}^3, \bar B_{\tau})
\begin{pmatrix}
\bar D& \bar l  \\
\bar l &\bar E\\
\end{pmatrix}
\begin{pmatrix}
\bar A_{\tau}^3\\
\bar B_{\tau}
\end{pmatrix}
+
\half h \bar F h +\half G^0 \bar G G^0.
\label{shiki99}
\end{eqnarray}
where
\begin{eqnarray}
\bar A&=&\del_i^2 +\del_{\tau}^2 -(2\pi T)^2 (2\varphi)^2-{g^2\over 4}v^2,
\nonumber\\
\bar B&=&\del_i^2 +\del_{\tau}^2 -(2\pi T)^2 \left(2\varphi\right)^2+\mu^2
-{\lambda\over 2}v^2 -{g^2\over 4}v^2~~=\bar C, 
\nonumber\\
%
%
\bar D&=&\del_i^2 +\del_{\tau}^2-{g^2\over 4}v^2, 
\nonumber\\
\bar E&=&\del_i^2 +\del_{\tau}^2-{g_Y^2\over 4}v^2, 
\nonumber\\
\bar F&=&\del_i^2 +\del_{\tau}^2+\mu^2 - {3\over 2}\lambda v^2, 
\nonumber\\
\bar G&=&\del_i^2 +\del_{\tau}^2 +\mu^2-{\lambda v^2\over 2}-{g^2+g_Y^2\over 4}v^2,
\nonumber\\
a&=&-2(2\pi T)(2\varphi)\del_{\tau},\quad
\bar g=-2(2\pi T)\left(2\varphi\right)\del_{\tau}, \quad \bar l={gg_Y\over 4}v^2 .
\label{shiki100}
\end{eqnarray}
The $\bar A_{\tau}^{1,2 }$ and $g^{1, 2}$  sectors are diagonalized by 
the original base defined in Eqs. (\ref{shiki67}) and (\ref{shiki70}). The 
$\bar A_{\tau}^3$ and $\bar B_{\tau}$ sector is diagonalized by the rotation matrix
given by Eq (\ref{shiki96}),
\begin{equation}
\begin{pmatrix}
Z_{\tau}\\
A_{\tau}^{\gamma}
\end{pmatrix}
=\begin{pmatrix}
c_w& -s_w\\
s_w &c_w
\end{pmatrix}
\begin{pmatrix}
\bar A_{\tau}^3\\
\bar B_{\tau}
\end{pmatrix}.
\label{shiki101}
\end{equation}
Then the eigenvalues in the momentum space are
\begin{eqnarray}
-k_i^2 -(2\pi T)^2(n-2\varphi)^2 -{g^2\over 4}v^2&\cdots &W_{\tau}^{\pm},\nonumber\\
-k_i^2 -(2\pi T)^2 n^2 -{g^2+g_Y^2\over 4}v^2 &\cdots &Z_{\tau},\nonumber\\
-k_i^2 -(2\pi T)^2n^2&\cdots &A_{\tau}^{\gamma},\nonumber\\
-k_i^2 -(2\pi T)^2(n-2\varphi)^2 +\mu^2 -{\lambda \over 2}v^2-{g^2\over 4}v^2&\cdots &G^{\pm},\nonumber\\
-k_i^2 -(2\pi T)^2n^2 +\mu^2 -{\lambda \over 2}v^2-{g^2+g_Y^2\over 4}v^2&\cdots &G^0,\nonumber\\
-k_i^2 -(2\pi T)^2n^2 +\mu^2 -{3\lambda \over 2}v^2&\cdots &h .
\label{shiki102}
\end{eqnarray}
Let us note that in the above calculations the terms proportional to  
$-\varphi + \theta/2$ or $p$ vanish and $\varphi +\theta/2$ becomes $2\varphi$ due 
to the ansatz.

The quadratic terms for the ghost sector is 
\begin{eqnarray}
i(\bar C^+,~ \bar C^-,~\bar C^3,~\bar C)
M_{\rm ghost}^2\bigl|_{\rm ansatz}
\begin{pmatrix}
C^-\\
C^+\\
C^3\\
C 
\end{pmatrix}
&=&
-i\bar C^+ \bar D_{W^{\pm}} C^-
-i\bar C^- \bar D_{W^{\pm}} C^+
\nonumber\\
&-&i
(\bar C^3, \bar C)
\begin{pmatrix}
\bar D^{A^3}& {gg_Y\over 4}v^2 \\
{gg_Y\over 4}v^2 & \bar D^B\\
\end{pmatrix}
\begin{pmatrix}
C^3\\
C 
\end{pmatrix}.
\label{shiki103}
\end{eqnarray}
where
\begin{eqnarray}
\bar D^{W^{\pm}}&=&\del_i^2+\Bigl(\del_{\tau}-i(2\pi T)(2\varphi)\Bigr)^2- {g^2\over 4}v^2,
\nonumber\\
\bar D^{A^3}&=&\del_i^2+\del_{\tau}^2- {g^2\over 4}v^2,\quad 
\bar D^B=\del_i^2+\del_{\tau}^2-{g_Y^2\over 4}v^2.
\label{shiki104}
\end{eqnarray}
By introducing the new bases by the rotation matrix (\ref{shiki96}), 
\begin{equation}
\begin{pmatrix}
\bar C_Z\\
\bar C_{\gamma}
\end{pmatrix}
=\begin{pmatrix}
c_w& -s_w\\
s_w &c_w
\end{pmatrix}
\begin{pmatrix}
\bar C^3\\
\bar C
\end{pmatrix},\quad
\begin{pmatrix}
C_Z\\
C_{\gamma}
\end{pmatrix}
=\begin{pmatrix}
c_w& -s_w\\
s_w &c_w
\end{pmatrix}
\begin{pmatrix}
C^3\\
C
\end{pmatrix},
\label{shiki105}
\end{equation}
the ghost sector is diagonalized.
The eigenvalues are given, in the momentum space, by
\begin{eqnarray}
-k_i^2-(2\pi T)^2(n-2\varphi)^2-{g^2\over 4}v^2&\cdots &\bar C^{\pm}, C^{\pm},
\nonumber\\
-k_i^2-(2\pi T)^2n^2-{g^2+g_Y^2\over 4}v^2&\cdots& \bar C_Z, C_Z,
\nonumber\\
-k_i^2-(2\pi T)^2n^2&\cdots &\bar C_{\gamma}, C_{\gamma}.
\label{shiki106}
\end{eqnarray}

Let us finally consider the fermion sector whose matrices under the ansatz are given by
\begin{eqnarray}
&&(\bar t_L, \bar t_R, \bar b_L, \bar b_R)M_{\rm quark}\big|_{\rm ansatz}
\begin{pmatrix}
t_L\\
t_R\\
b_L\\
b_R
\end{pmatrix}
+({\bar\nu}_L, \bar \tau_L, \bar \tau_R,)M_{\rm lepton}\big|_{\rm ansatz}
\begin{pmatrix}
\nu_L\\
\tau_L\\
\tau_R
\end{pmatrix}
\label{shiki107}
\\
&=&
(\bar t_L, \bar t_R)
\begin{pmatrix}
-i\bar D_{t_L}& {f_t\over\sqrt{2}}v \\
{f_t\over\sqrt{2}}v  & -i\bar D_{t_R}\\
\end{pmatrix}
\begin{pmatrix}
t_L\\
t_R
\end{pmatrix}
+
(\bar b_L, \bar b_R)
\begin{pmatrix}
-i\bar D_{b_L}& {f_b\over\sqrt{2}}v \\
{f_b\over\sqrt{2}}v  & -i\bar D_{b_R}\\
\end{pmatrix}
\begin{pmatrix}
b_L\\
b_R
\end{pmatrix}\nonumber\\
&+&
(\bar \tau_L, \bar \tau_R)
\begin{pmatrix}
-i\bar D_{{\tau}_L}& {f_{\tau}\over\sqrt{2}}v \\
{f_{\tau}\over\sqrt{2}}v  & -i\bar D_{{\tau}_R}\\
\end{pmatrix}
\begin{pmatrix}
\tau_L\\
\tau_R
\end{pmatrix}
-i\bar \nu_L \bar D_{\nu_L}\nu_L ,
\label{shiuki108}
\end{eqnarray}
where
\begin{eqnarray}
\bar D_{t_L}&\equiv &\gamma_{\tau}\Bigl(\del_{\tau}-i(2\pi T)
\bigl(\omega_r+ {4\over 3}\varphi\bigr)\Bigr)+\gamma_i\del_i ~~=\bar D_{t_R},\nonumber\\
\bar D_{b_L}&\equiv &\gamma_{\tau}\Bigl(\del_{\tau}-i(2\pi T)
\bigl(\omega_r -{2\over 3}\varphi\bigr)\Bigr)+\gamma_i\del_i~~=\bar D_{b_R},\nonumber\\
\bar D_{\tau_L}&\equiv &\gamma_{\tau}\Bigl(\del_{\tau}+i(2\pi T)2\varphi\Bigr)
+\gamma_i\del_i~~=\bar D_{\tau_R},\nonumber\\
\bar D_{\nu_L}&\equiv &\gamma_{\tau}\del_{\tau}+\gamma_i\del_i.
\label{shiki109}
\end{eqnarray}
Let us note that thanks to the ansatz the diagonal components of each matrix 
for the top, bottom and tau sectors
become identical. The eigenvalues in the momentum space are
\begin{eqnarray}
k_i^2 + (2\pi T)^2\left(n+\half -\omega_r -{4\over 3}\varphi\right)^2+{f_t^2\over 2}v^2 &\cdots& t_L, t_R, 
\nonumber\\
k_i^2 + (2\pi T)^2\left(n+\half -\omega_r +{2\over 3}\varphi\right)^2+{f_b^2\over 2}v^2&\cdots&b_L, b_R,
\nonumber \\
k_i^2 + (2\pi T)^2\left(n+\half +2\varphi\right)^2+{f_{\tau}^2\over 2}v^2&\cdots&\tau_L, \tau_R, 
\nonumber\\
k_i^2 + (2\pi T)^2\left(n+\half\right)^2&\cdots& \nu_L .
\label{shiki110}
\end{eqnarray}
The half-integer in the Matsubara mode $n$ is due to the Fermi statistics for fermions.
Since the quarks have the color degrees of freedom, the eigenvalues also depend on the order 
parameter $\omega_r$ of the vacuum expectation value $\vev{G_{\tau}}$.

Taking account of the eigenvalues obtained above, the one-loop contributions
to the effective potential are given by
\begin{eqnarray}
V^{\rm one-loop}&=&
(6+2-4)\times {1\over 2i}\int_k {\rm ln}\biggl[k_i^2 +(2\pi T)^2(n-2\varphi)^2 +{g^2\over 4}v^2\biggr]
\nonumber\\
&+&
(3+1-2)\times {1\over 2i}\int_k {\rm ln}\biggl[k_i^2 +(2\pi T)^2n^2 +{g^2+g_Y^2\over 4}v^2\biggr]
\nonumber\\
&+&
(3+1-2)\times {1\over 2i}\int_k {\rm ln}\biggl[k_i^2 +(2\pi T)^2n^2 \biggr]
\nonumber\\
&+&
(4-2)\sum_{r, q=1}^3{1\over 2i}\int_k{\rm ln}
\biggl[k_i^2 +(2\pi T)^2(n+\omega_r - \omega_q)^2\biggr]
\nonumber\\
&+&
2\times {1\over 2i}\int_k {\rm ln}\biggl[k_i^2 +(2\pi T)^2(n-2\varphi)^2-\mu^2 +{\lambda\over 2}v^2 
+{g^2\over 4}v^2\biggr]
\nonumber\\
&+&
1\times {1\over 2i}\int_k {\rm ln}\biggl[k_i^2 +(2\pi T)^2n^2 -\mu^2
+{\lambda\over 2}v^2+{g^2+g_Y^2\over 4}v^2\biggr]
\nonumber\\
&+&
1\times {1\over 2i}\int_k {\rm ln}\biggl[k_i^2 +(2\pi T)^2n^2 -\mu^2
+{3\lambda\over 2}v^2\biggr]
\nonumber\\
&+&
(-1)2^2 \sum_{r=1}^3 {1\over 2i}\int_k {\rm ln}\biggl[k_i^2 +(2\pi T)^2
\Bigl(n+\half-\omega_r -{4\over 3}\varphi\Bigr)^2 
+{f_t^2\over 2}v^2\biggr]\nonumber\\
&+&
(-1)2^2 \sum_{r=1}^3 {1\over 2i}\int_k {\rm ln}\biggl[k_i^2 +(2\pi T)^2
\Bigl(n+\half-\omega_r +{2\over 3}\varphi\Bigr)^2  
+{f_b^2\over 2}v^2\biggr]\nonumber\\
&+&
(-1)2^2 \times {1\over 2i}\int_k {\rm ln}\biggl[k_i^2 +(2\pi T)^2\Bigl(n+\half +2\varphi\Bigr)^2  
+{f_{\tau}^2\over 2}v^2\biggr]\nonumber\\
&+&
(-1){2^2\over 2} \times {1\over 2i}\int_k {\rm ln}\biggl[k_i^2 +(2\pi T)^2\Bigl(n+\half\Bigr )^2 \biggr],
\label{shiki111}
\end{eqnarray}
where we have defined 
\begin{equation}
\int_k \equiv iT \sum_{n=-\infty}^{\infty}\int {d^3 k\over(2\pi)^3}.
\label{shiki112}
\end{equation}
The first to fourth lines come from 
$W_{i, \tau}^{\pm}, Z_{i, \tau}, A^{\gamma}_{i,\tau}$ and $G_{i,\tau}^{\alpha}$, respectively
together with the ghost fields $C^{\pm}, \bar C^{\pm}, C_Z, \bar C_Z, C_{\gamma}, 
\bar C_{\gamma}, C_s^{\alpha}, \bar C_s^{\alpha}$. The fields $G^{\pm}, G^0$ and $h$ contribute 
to the fifth, sixth and seventh lines, respectively. The last four lines are the fermion contributions. In 
addition to the usual order parameter $v$, the one-loop contributions depend on the new 
order parameters $\varphi$ and $\omega_r$. 
%
%
%
As discussed in the text, the effective potential at the one-loop level
consists of the zero and finite temperature part,
\begin{equation}
V^{\rm one-loop}=V^{T=0} + V^{T\neq 0}, 
\label{shiki114}
\end{equation}
where 
\begin{eqnarray}
&&V^{T=0}\nonumber\\
&=& -{\mu^2\over 2}v^2+{\lambda\over 8}v^4\nonumber\\
&+&{4\over 4(4\pi)^2}m_W(v)^4\biggl({\rm ln}{m_W(v)^2\over M^2}-{3\over 2}\biggr)
+{2\over 4(4\pi)^2}m_Z(v)^4\biggl({\rm ln}{m_Z(v)^2\over M^2}-{3\over 2}\biggr)
\nonumber\\
&+&{2\over 4(4\pi)^2}m_{G^{\pm}}(v)^4
\biggl({\rm ln}{m_{G^{\pm}}(v)^2\over M^2}-{3\over 2}\biggr)
+{1\over 4(4\pi)^2}m_{G^{0}}(v)^4
\biggl({\rm ln}{m_{G^{0}}(v)^2\over M^2}-{3\over 2}\biggr)
\nonumber\\
&+&{1\over 4(4\pi)^2}m_{h}(v)^4
\biggl({\rm ln}{m_{h}(v)^2\over M^2}-{3\over 2}\biggr)
-{12\over 4(4\pi)^2}m_{t}(v)^4
\biggl({\rm ln}{m_{t}(v)^2\over M^2}-{3\over 2}\biggr)
\nonumber\\
&-&{12\over 4(4\pi)^2}m_{b}(v)^4
\biggl({\rm ln}{m_{b}(v)^2\over M^2}-{3\over 2}\biggr)
-{4\over 4(4\pi)^2}m_{\tau}(v)^4
\biggl({\rm ln}{m_{\tau}(v)^2\over M^2}-{3\over 2}\biggr)
\label{shiki115}
\end{eqnarray}
and
\begin{eqnarray}
&&V^{T\neq 0}\nonumber\\
&=&-4{2\over (2\pi)^2}T^4\sum_{m=1}^{\infty}
{1\over m^4}\cos[2\pi m(2\varphi)]\left({m_W(v)^2\over T^2}m^2\right)K_2\left({m_W(v)\over T}m\right)
\nonumber\\
&-&2{2\over (2\pi)^2}T^4\sum_{m=1}^{\infty}
{1\over m^4}\left({m_Z(v)^2\over T^2}m^2\right)K_2\left({m_Z(v)\over T}m\right)
\nonumber\\
&-&2{2\over (2\pi)^2}T^4\sum_{m=1}^{\infty}
{1\over m^4}\cos[2\pi m(2\varphi)]
\left({m_{G^{\pm}}(v)^2\over T^2}m^2\right)K_2\left({m_{G^{\pm}}(v)\over T}m\right)
\nonumber\\
&-&1{2\over (2\pi)^2}T^4\sum_{m=1}^{\infty}
{1\over m^4}
\left({m_{G^{0}}(v)^2\over T^2}m^2\right)K_2\left({m_{G^{0}}(v)\over T}m\right)
\nonumber\\
&-&1{2\over (2\pi)^2}T^4\sum_{m=1}^{\infty}
{1\over m^4}
\left({m_{h}(v)^2\over T^2}m^2\right)K_2\left({m_{h}(v)\over T}m\right)
\nonumber\\
&+&4{2\over (2\pi)^2}T^4\sum_{r=1}^3\sum_{m=1}^{\infty}
{(-1)^m\over m^4}\cos\Bigl[2\pi m\bigr(\omega_r + {4\over 3}\varphi\bigl)\Bigr]
\left({m_{t}(v)^2\over T^2}m^2\right)K_2\left({m_{t}(v)\over T}m\right)
\nonumber\\
&+&4{2\over (2\pi)^2}T^4\sum_{r=1}^3\sum_{m=1}^{\infty}
{(-1)^m\over m^4}\cos\Bigl[2\pi m\bigr(\omega_r -{2\over 3}\varphi\bigl)\Bigr]
\left({m_{b}(v)^2\over T^2}m^2\right)K_2\left({m_{b}(v)\over T}m\right)
\nonumber\\
&+&4{2\over (2\pi)^2}T^4\sum_{m=1}^{\infty}
{(-1)^m\over m^4}\cos[2\pi m(2\varphi)]
\left({m_{\tau}(v)^2\over T^2}m^2\right)K_2\left({m_{\tau}(v)\over T}m\right)
\nonumber\\
&-&2{2\over(2\pi)^2}T^4\sum_{r, q=1}^3\sum_{m=1}^{\infty}{2\over m^4}\cos[2\pi m(\omega_r - \omega_q)].
\label{shiki116}
\end{eqnarray}
The last line in Eq. (\ref{shiki116}) comes from the $SU(3)_c$ gauge sector. The 
new order parameters $\varphi$ and $\omega_r$ enter into the finite temperature part of the one-loop 
effective potential. Here we have defined the notations,
\begin{eqnarray}
&&m_W(v)^2={g^2\over 4}v^2,~~m_Z(v)^2={g^2+g_Y^2\over4}v^2,~~m_h(v)^2=-\mu^2+{3\lambda\over 2} v^2,
\nonumber\\
&&m_{G^{\pm}}(v)^2=-\mu^2+ {\lambda\over 2} v^2+m_W(v)^2,~~
m_{G^0}(v)^2=-\mu^2+ {\lambda\over 2} v^2+m_Z(v)^2,\nonumber\\
&&m_{t}(v)^2={f_t^2\over 2}v^2,~~m_{b}(v)^2={f_b^2\over 2}v^2,~~m_{\tau}(v)^2={f_{\tau}^2\over 2}v^2.
\label{shiki117}
\end{eqnarray}
$K_2(z)$ is the modified Bessel function defined in Eq. (\ref{shiki35}).
%
%


\end{document}